\documentclass[fleqn,usenatbib]{mnras}

\usepackage{newtxtext,newtxmath}
\usepackage[T1]{fontenc}

\DeclareRobustCommand{\VAN}[3]{#2}
\let\VANthebibliography\thebibliography
\def\thebibliography{\DeclareRobustCommand{\VAN}[3]{##3}\VANthebibliography}

\usepackage{graphicx}	% Including figure files
\usepackage{amsmath}	% Advanced maths commands
\usepackage{physics}
\usepackage{mhchem}
\title[Chemistry in Molecular Cloud Formation]{Chemical Evolution during Molecular Cloud Formation Triggered by an Interstellar Shock Wave: Dependence on Shock Parameters and Comparison with Molecular Absorption Lines}

\author[Y. Komichi et al.]{
Yuto Komichi,$^{1}$\thanks{E-mail: komichi@astron.s.u-tokyo.ac.jp}
Yuri Aikawa,$^{1}$
Kazunari Iwasaki$^{2}$
and Kenji Furuya$^{1}$
\\
$^{1}$Department of Astronomy, Graduate School of Science, The University of Tokyo, 7-3-1, Hongo, Bunkyo-ku, Tokyo, 113-0033, Japan\\
$^{2}$Center for Computational Astrophysics, National Astronomical Observatory of Japan, 2-21-1, Osawa, Mitaka, Tokyo, 181-8588, Japan
}

\date{Accepted XXX. Received YYY; in original form ZZZ}

\pubyear{\the\year{}}

\begin{document}
\label{firstpage}
\pagerange{\pageref{firstpage}--\pageref{lastpage}}
\maketitle

\begin{abstract}
We investigate chemistry in the compression layer behind the interstellar shock waves, where molecular cloud formation starts. We perform three-dimensional magnetohydrodynamics simulations of converging flows of atomic gas \textrm{with shock parameters of inclination between the interstellar magnetic field and the shock wave, pre-shock density, and shock velocity. Then we derive 1D mean-flow models, along which we calculate a detailed gas-grain chemical reaction network as a post process with various chemical parameters, i.e. cosmic-ray ionization rate, abundances of PAHs, and metals in the gas phase.} While carbon chains reach their peak abundances when atomic carbon is dominant in the pseudo-time-dependent models of molecular clouds, such behavior is less significant in our models since the visual extinction of the compression layer is low ($\la 1$ mag) when \textrm{atomic carbon} is abundant. Carbon chains, CN, and HCN increase at $A_V \ga 1$ mag, where the gas-phase C/O ratio increases due to water ice formation. Shock parameters affect the physical structure and the evolutional timescale of the compression layer, and thus molecular evolution. Carbon chains are more abundant in models with higher post-shock density and slower gas accumulation. We calculate molecular column densities in the compression layer and compare them with the observations of diffuse and translucent clouds, which show reasonable agreement for water ice, carbon chains, and HCO$^+$. The observed variation of their column densities could be due to the difference in shock parameters and chemical parameters. The column density of CN is overestimated, for which we discuss possible reasons.
\end{abstract}


%%%%%%%%%%%%%%%%%%%%%%%%%%%%%%%%%%%%%%%%%%%%%%%%%%

%%%%%%%%%%%%%%%%% BODY OF PAPER %%%%%%%%%%%%%%%%%%

\section{Introduction}

\label{sec:intro}

Molecular cloud formation is a pivotal astrophysical process that sets the initial conditions of star formation.
Observational and theoretical work in recent years shows that the molecular cloud formation is triggered by interstellar shock waves \citep{2015A&A...580A..49I,2023ASPC..534..233P}, which are driven by e.g. supernovae \citep{1974ApJ...188..501C,1977ApJ...218..148M,1988ApJ...334..252C} and HII regions \citep{1978ppim.book.....S}. Expanding hot gas sweeps up and compresses the HI gas in galactic disks, making dense cold gas behind the shock wave. 
Magnetohydrodynamic (MHD) simulations show that shock waves can also induce filamentary structures characteristic of molecular clouds \citep{2013ApJ...774L..31I,2018PASJ...70S..53I,2021ApJ...916...83A}. Observations also indicate the relationship between filament formation and shock waves \citep[e.g.][]{2018PASJ...70...96A}.

The cloud formation process is important not only for astrophysics but also for astrochemistry. To date, more than 200 interstellar molecules have been detected \citep{McGuire_2022}.
Although there are many studies on the chemical modeling of interstellar clouds, most of them calculate the chemical abundance of molecular clouds as a function of time assuming a static and uniform physical condition and that H is already converted to H$_2$, i.e. pseudo-time-dependent model \citep{1980ApJ...239..151P,2019ApJ...871..238H}.
Some pioneering work studied the chemical evolution of a molecular cloud in its formation stage from HI gas with time-dependent physical parameters \citep{2004ApJ...612..921B,2010A&A...515A..66H,2015A&A...584A.124F}. They used one-dimensional (1D) steady-state shock models and solved the chemical evolution in the post-shock gas as a function of time, which is converted to visual extinction. Their models succeeded in providing the basic chemical properties regarding main gaseous species such as H$_2$ and CO \citep{2004ApJ...612..921B}, ice species on grain surface \citep{2010A&A...515A..66H}, and water deuteration \citep{2015A&A...584A.124F}. Since H$_2$ does not emit lines in cold ($\la$ 100 K) regions, theoretical studies on major molecular abundances such as atomic C and CO relative to H$_2$ are essential for observational studies of cloud formation. Chemistry in interstellar clouds is mostly non-equilibrium, and thus the astrochemistry of various molecules, ices, and isotope ratios could also be used to validate the cloud formation hypothesis via comparison with observations.

While these previous studies adopted a 1D shock model, MHD simulations in recent years show that multi-dimensional structures with magnetic fields are essential in molecular cloud formation.
Without the magnetic field, molecular clouds can be directly formed by a single shock compression, even from pure warm neutral medium (WNM) \citep{2002ApJ...564L..97K,2006ApJ...648.1052H,Vazquez-Semadeni_2006,2007A&A...465..431H}. On the other hand, with the interstellar magnetic fields with typical strength of several $\mathrm{\mu G}$, the magnetic pressure can prevent atomic gas from evolving into molecular clouds and the gas is converted to cold neutral medium (CNM) \textrm{\citep{2008ApJ...687..303I,Inoue_2009,2009ApJ...695..248H,2015MNRAS.451.3340K,2019ApJ...873....6I,Iwasaki_2022}}. \citet{2019ApJ...873....6I} show that there is a critical angle for efficient compression (i.e., achieving high post-shock density), which is determined by the balance between the turbulence due to instability behind shock waves and the magnetic pressure in the compression layers. The inclination of the interstellar magnetic field against the shock wave directly affects the evolution of compression layers or the initial phase of molecular cloud formation.  

In the present work, we investigate the chemical evolution of atomic gas to molecular gas based on the 3D MHD simulation of interstellar shock compression.
We aim to adopt detailed chemical models that include thousands of chemical reactions to investigate the dependence of various molecular abundances in the gas phase and ice mantles on the shock parameters. However, it is computationally too expensive to directly couple such a detailed chemical network with the multi-dimensional MHD simulations. 
We thus construct \textrm{1D} models of the physical evolution of interstellar medium (ISM) from the results of the \textrm{3D} MHD simulations and conduct the network calculation as a post process. 
Specifically, we model the mean evolution of the compression layers based on the 3D MHD simulations and conduct the chemical network calculations along the flow. This is an extension of the previous chemical models based on 1D shock and is the first step in studying the detailed chemical evolution of the compression layers taking into account the effect of multi-dimensional flow.

\textrm{In both the original 3D simulations and the derived 1D models, the column density and thus the visual extinction $A_V$ of the compression layer increases with time.
In the present work, we focus on the early phase of molecular cloud formation and conduct a series of calculations until $A_V$ reaches $\sim 5$ mag. Typical hydrogen number density in the compression layer is $10^2-10^3\,\mathrm{cm^{-3}}$.} We thus compare our results with observations of diffuse and translucent clouds \citep{2006ARA&A..44..367S}.
The chemical characteristics of diffuse clouds have been studied by absorption lines of molecules observed toward various kinds of light sources such as stars, massive star-forming regions, and extragalactic sources \citep[e.g.][]{1996A&A...307..237L,2002A&A...384.1054L,2006A&A...448..253L,2008ApJ...687.1075S,2010A&A...518L.110G,2010A&A...520A..20G,2014A&A...564A..64L,2022ApJ...928...79R,2023A&A...670A.111K}. Such observations reveal statistics of the chemical properties of diffuse clouds in our galaxy. For example, while the column densities of different kinds of interstellar molecules are well correlated, they also show a scatter of orders of magnitude \citep[e.g. see Figure 4. to 6. in][]{2018A&A...610A..43R}. The origin of this scatter is still not well understood. We compare our results with these observations to investigate if the scatter can be explained by the chemistry with different shock models and/or other parameters such as cosmic-ray ionization rate.

\textrm{Even in such an early phase of cloud evolution, it is important to consider not only gas-phase reactions but also grain-surface reactions. For example, \citet{2015A&A...584A.124F} showed that water ice becomes one of the major oxygen reservoirs at $A_V\sim 1$ mag. It is well known that water ice formation enhances the elemental C/O ratio in the gas phase, which significantly affects carbon chemistry \citep[e.g.][]{Pratap_1997,Terzieva_1998, Aikawa_2003, Ruaud_2018}. If the C/O ratio is $< 1$ and CO formation is efficient, almost all carbon is used to form CO and the formation of other C-bearing molecules is suppressed. On the other hand, in the case of C/O > 1, even if all oxygen is used for the CO formation, there is leftover carbon, which can be used to form carbon chains and CN-bearing molecules. The water formation starts from the adsorption of O atoms on to grain surfaces, where they are hydrogenated. Since the adsorption timescale is comparable to the cloud formation timescale $\sim 60\, \mathrm{Myr} (A_V/1\,\mathrm{mag})(v_\mathrm{flow}/10\,\mathrm{km/s})^{-1}(n/1\,\mathrm{cm^{-3}})^{-1}$ \citep{Inoue_2012}, where $v_\mathrm{flow}$ and $n$ are the inflowing gas velocity and density, respectively, the C/O ratio in the gas phase also depends on the physical evolution of molecular clouds. We look into these properties using detailed chemical network calculations.}

The rest of the paper is organized as follows. The details of the shock model are summarized in Section \ref{sec:phys}. The method of chemical network calculations is described in Section \ref{sec:chem}. The results of the network calculation including a parameter survey are shown in Section \ref{sec:result}. The comparison between our results and observations of diffuse clouds and other discussions are presented in Section \ref{sec:discussion}. Finally, we summarize our results in Section \ref{sec:conclusion}.

\section{Physical Models} \label{sec:phys}

In this section, we describe the setup of our MHD simulations and the methodology for modeling the mean evolution of the compression layers along the shock propagation, which are used for chemical network calculations.

\subsection{MHD simulation}

\subsubsection{Basic Equations}
\label{sec:basicequ}

The basic equations of MHD are given as follows:
\begin{equation}
    \pdv{\rho}{t}+\nabla\cdot (\rho \vb*{u})=0,
\end{equation}
\begin{equation}
        \pdv{t}(\rho\vb*{u})+\boldsymbol\nabla\cdot\biggl[\rho\vb*{u}\boldsymbol\otimes\vb*{u}+\left(p+\frac{\vb*{B}^2}{8\pi}\right)\vb*{I}-\frac{\vb*{B}\boldsymbol\otimes\vb*{B}}{4\pi}\biggr]=0,
\end{equation}
\begin{equation}
        \pdv{e}{t}+\boldsymbol\nabla\cdot\biggl[\left(e+p+\frac{\vb*{B}^2}{8\pi}\right)\vb*{u}-\frac{\vb*{B}(\vb*{u}\cdot\vb*{B})}{4\pi}-\kappa\boldsymbol\nabla T\biggr]=\mathcal{G},
\end{equation}
and
\begin{equation}
    \pdv{\vb*{B}}{t}-\boldsymbol\nabla\boldsymbol\cross(\vb*{u}\boldsymbol\cross\vb*{B})=0.
\end{equation}
Here, $\rho$ is the mass density, $\vb*{u}$ is the velocity, $p$ is the thermal pressure, $\vb*{B}$ is the magnetic field, $e$ is the total energy density, $T$ is the temperature, $\kappa$ is the thermal conductivity, and $\mathcal{G}$ is the net heating rate per unit volume. The thermal conductivity of hydrogen atoms is set to be $\kappa=2.5\times 10^3\,\sqrt{T}\,\mathrm{erg\,cm^{-1}\,s^{-1}\,K^{-1}}$ \citep{1953ApJ...117..431P}. We refer to \citet{2002ApJ...564L..97K} for the heating and cooling functions ($\Gamma, \Lambda$), and updated the heating function as follows, including dust extinction and cosmic-ray heating, 
\begin{equation}
    \mathcal{G}=n_\mathrm{H}\Gamma - n_\mathrm{H}^2\Lambda,
\end{equation}
\begin{equation} \label{equ:heating}
    \begin{split}
        \Gamma =& 2.0\times 10^{-26}G_0\exp(-\alpha A_V)\\&+3.2\times 10^{-11}\zeta_\mathrm{CR}\,\mathrm{erg\,s^{-1}},
    \end{split}
\end{equation}
where $n_\mathrm{H}=\rho/1.4m_\mathrm{H}$ is the hydrogen nucleus number density, $G_0$ is the strength of interstellar radiation field in the Habing unit \textrm{\citep{1968BAN....19..421H}}, $A_V$ is the visual extinction, $\alpha=2.5$ is the coefficient for dust extinction \citep{2004ApJ...612..921B}, and $\zeta_\mathrm{CR}$ is the cosmic-ray ionization rate. The cosmic-ray heating rate (the second term in equation (\ref{equ:heating})) \textrm{and $\zeta_\mathrm{CR}$ of $1\times 10^{-17}\,\mathrm{s^{-1}}$ are} taken from \citet{1978ApJ...222..881G}. We mainly focus on the early phase of cloud formation (\textrm{e.g. until $A_V$ of the compression layer reaches 5 mag}) and neglect self-gravity in our MHD simulations.

\textrm{
We set $G_0=1$ in the present work since we are interested in a typical environment of molecular cloud formation in the solar neighborhood. The gas temperature depends on $G_0$, but at the gas densities of $\ge 100$ cm$^{-3}$, which corresponds to the density in the compression layer, the dependence is rather weak; e.g. the gas temperature is increased only by a factor of a few  with $G_0=10$ compared with the case of $G_0=1$. According to \citet{2017A&A...604A..58H}, the dependence of dust temperature on $G_0$ is even weaker; the dust temperature is proportional to $G_0^{1/6}$. Therefore, the effect of $G_0$ on our physical models would be negligible, as long as we are interested in interstellar gas outside high-mass star-forming regions.}

\subsubsection{Calculation Setup}

\begin{figure}
\includegraphics[width=\columnwidth]{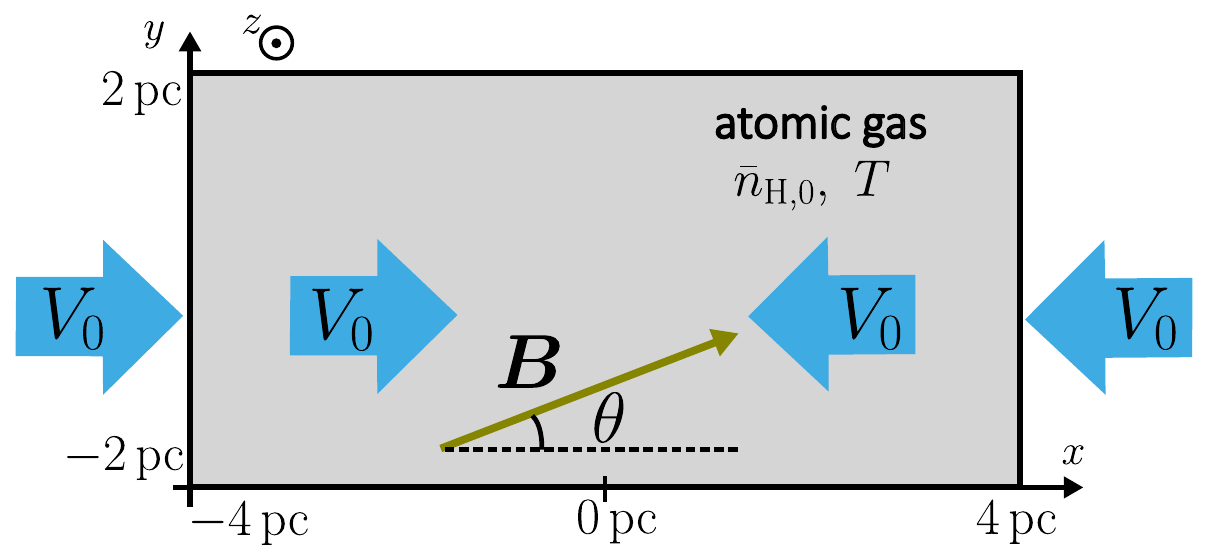}
\caption{The schematic picture of initial and boundary conditions of our MHD simulations.
\label{fig:mhd_setting}}
\end{figure}

\begin{table}
	\centering
	\caption{Parameter sets used in our MHD simulations. $\bar{p}_\mathrm{0}$ is the equilibrium pressure at the mean hydrogen number density $\bar{n}_\mathrm{H,0}$, assuming $A_V=0$\,mag. $k_\mathrm{b}$ is Boltzmann's constant. We choose N10V10sin020 as a fiducial model.}
	\label{tab:Table1}
	\begin{tabular}{lcccc} % four columns, alignment for each
		\hline
		{Model Name} & {$\bar{n}_\mathrm{H,0}$}& {$V_0$} & {$\sin\theta$} & {$\bar{p}_\mathrm{0}/k_\mathrm{b}$}\\
         & {$(\mathrm{cm^{-3}})$}& {$(\mathrm{km\,s^{-1}})$} & {} & {$(K\,\mathrm{cm^{-3}})$}\\
		\hline
	   N10V10sin005    & 10 & 10 & 0.05 & 1400\\
          N10V10sin020    & 10 & 10 & 0.20 & 1400\\
          N10V10sin030    & 10 & 10 & 0.30 & 1400\\
          N10V10sin050    & 10 & 10 & 0.50 & 1400\\
          N10V10sin070    & 10 & 10 & 0.70 & 1400\\
          N10V10sin100    & 10 & 10 & 1.00 & 1400\\
          N10V30sin030    & 10 & 30 & 0.30 & 1400\\
          N02V50sin030     & 2  & 50 & 0.30 & 2800\\
		\hline
	\end{tabular}
\end{table}

\begin{figure}
\includegraphics[width=\columnwidth]{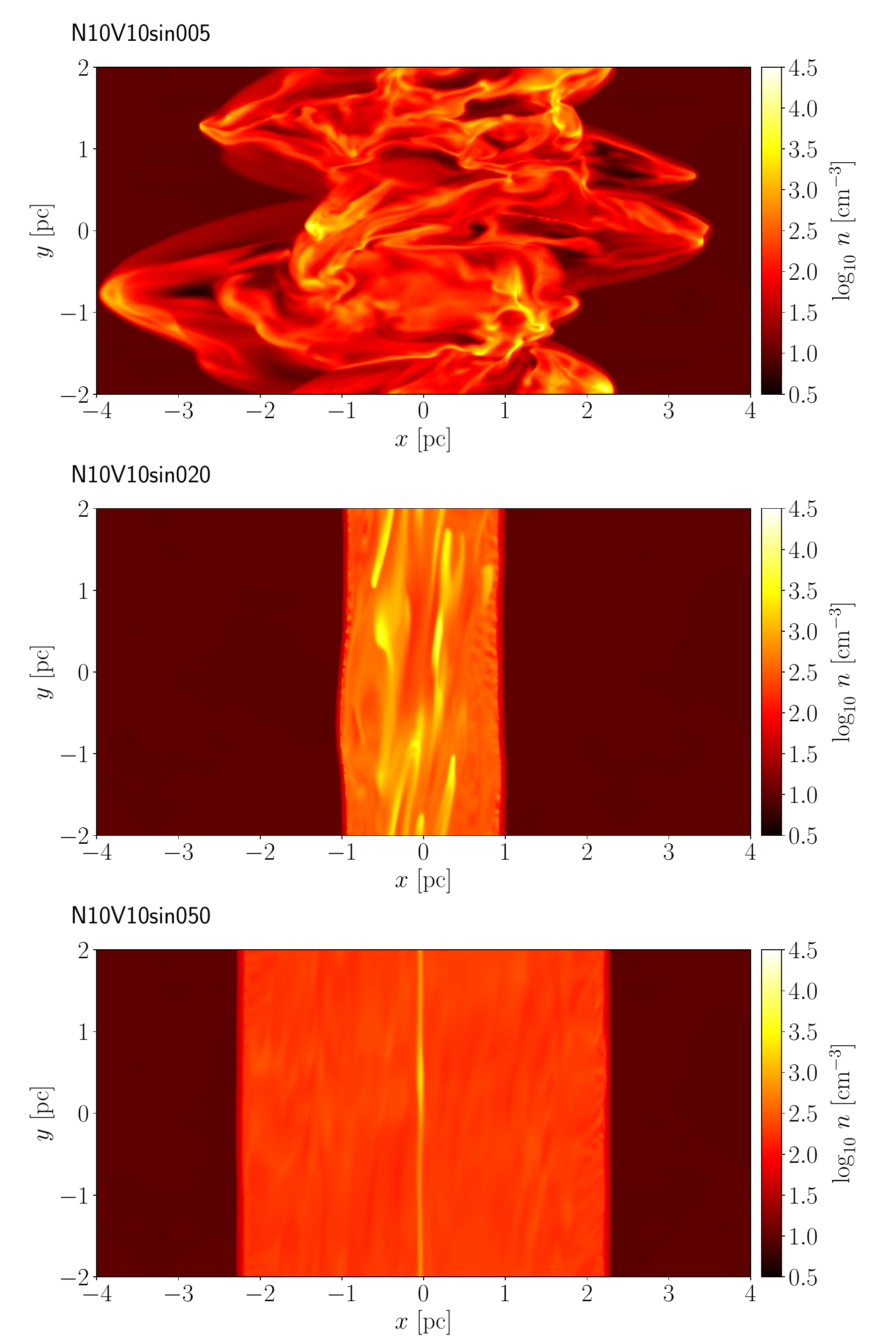}
\caption{The number density of hydrogen nuclei on the $z=0$ plane at $t=5\,$Myr in the models of N10V10sin005 (top), N10V10sin020 (middle), and N10V10sin050 (bottom).}
\label{fig:densityslice}
\end{figure}

\begin{figure*}
\includegraphics[width=2\columnwidth]{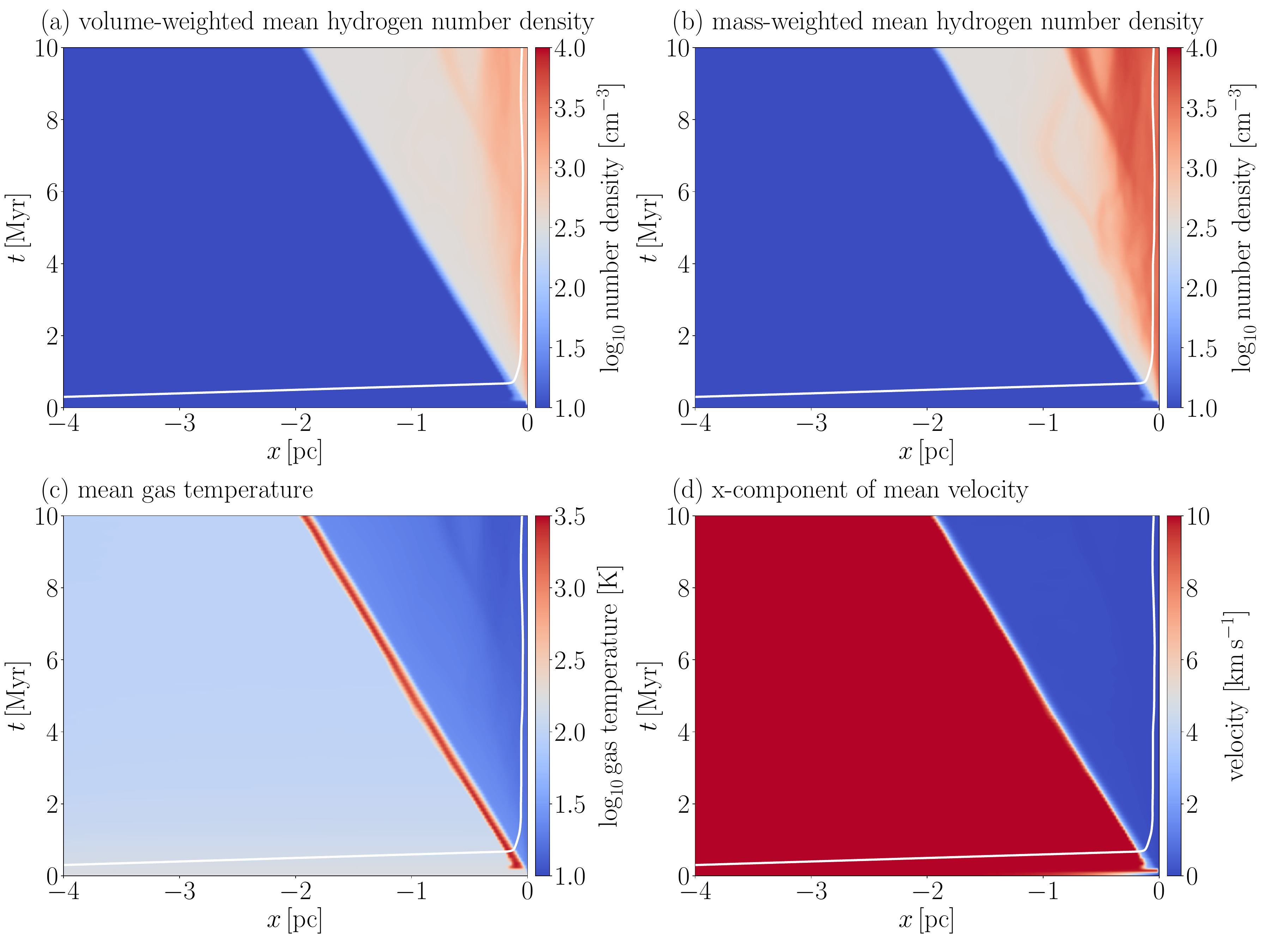}
\caption{Physical quantities in the fidcucial model (N10V10sin020) averaged over the $y-z$ plane at each position $x$ and time. Panels (a) - (d) show the volume-weighted mean hydrogen number density, the mass-weighted mean hydrogen number density, the mean gas temperature, and the $x$-component of mean velocity, respectively. The white line shows the trajectory of a particle that moves along the mean velocity (d) from the boundary.
}
\label{fig:analysis}
\end{figure*}

We simulate the interstellar shock of a conversing flow (Figure \ref{fig:mhd_setting}).
The size of our simulation box is $L_x=8\,\mathrm{pc}$, and $L_y=L_z=4\,\mathrm{pc}$. This calculation region is divided into $512 \times 256 \times 256$ cells with the spatial resolution of $\Delta x \simeq 0.02\,\mathrm{pc}$, so that the typical cooling length of WNM ($\sim$ a few $\mathrm{pc}$) is well-resolved \citep{2020ApJ...905...95K}.

Initially, the gas is completely atomic. The velocity field is directed to the $x=0$ plane in order to initiate shock compression at the center of our calculation box;
\begin{equation}
    u_x=-V_0\tanh{\left(\frac{x}{0.5\,\mathrm{pc}}\right)}.
\end{equation}
The initial magnetic field is on the $x-y$ plane; it is uniform and inclined from the $x$-axis with an angle $\theta$. The initial field strength is set to be $2\,\mathrm{\mu G}$. 
The averaged hydrogen number density of the initial atomic gas is $\bar{n}_\mathrm{H,0}$. We set the initial density dispersion by $P(\log n)\propto \vb*{k}^{-2}$ as the seed of turbulence, where $P(\log n)$ is the power spectrum and $\vb*{k}$ is the wave number and varies from $2\pi/L_y$ to $32\pi/L_y$. The mean density dispersion is set to be $1\%$. In reality, the density dispersion would be larger and the initial gas would be two-phase (i.e., WNM and CNM). 
We set the small density dispersion for the fiducial model since the main motivation of the present work is to study the chemical evolution of the mean flow. More realistic simulations with a higher dispersion are postponed to future studies.
The initial gas pressure is set to be uniform with a value given by the thermal equilibrium state at the averaged density $\bar{n}_\mathrm{H,0}$ and $A_V=0\,\mathrm{mag}$.
The boundary conditions of the $x$ direction are set so that the initial gas is continuously injected with velocity $V_0$, but the pressure of the injected gas is set to be the equilibrium value at $A_V=A_V^\mathrm{eff}$, which is given below. We adopt periodic conditions for $y$ and $z$ directions.

In order to calculate the local net heating rate (equation (\ref{equ:heating})), it is necessary to evaluate the visual extinction (or total hydrogen column density). In this study, we use a two-ray approximation. The effective visual extinction is calculated by integrating total hydrogen number density from the $+x$ and $-x$ boundaries: 
\begin{equation}
\begin{split}
    A_V^\mathrm{eff}(x,y,z) = -\frac{1}{\alpha}\ln\,&\biggl[\frac{1}{2}\exp(-\alpha A_V^+(x,y,z))\\
    &+\frac{1}{2}\exp(-\alpha A_V^-(x,y,z))\biggr],
\end{split}
\end{equation}
\begin{equation}
    A_V^+(x,y,z)=\frac{1}{N_\mathrm{H,0}}\int_{-L_x/2}^{x}\,(n_\mathrm{H}(x',y,z)-\bar{n}_{\mathrm{H,0}})\,\dd x',
\end{equation}
\begin{equation}
    A_V^-(x,y,z)=\frac{1}{N_\mathrm{H,0}}\int_{x}^{L_x/2}\,(n_\mathrm{H}(x',y,z)-\bar{n}_{\mathrm{H,0}})\,\dd x',
\end{equation}
where $N_\mathrm{H,0}=2.0\times 10^{21}\,\mathrm{cm^{-2}}$ corresponds to $A_V=1$ mag.
This method is justified because the shock-compressed region has a sheet-like structure. 
Here, we subtract the pre-shock density $\bar{n}_\mathrm{H,0}$ so that $A_V^\mathrm{eff}$ does not depend on the size of the simulation box (see also \ref{sec:analysis_flow_1dmodel}).

Our calculations are conducted with Athena++ \citep{2020ApJS..249....4S} and simulate the evolution up to 10 Myr or until the shock fronts reach the calculation boundary of the $x$ direction. The calculation time is $(1-5)\times 10^4$ core-hours on the Cray XC50 supercomputer, depending on initial parameters.

\subsubsection{Initial Parameters}
\label{sec:initialparameters}

According to the recent observations of HI clouds \citep[e.g.][]{2009ApJ...705..144F,2022ApJ...941...62H}, they typically have a number density larger than that of pure WNM $(\sim 1\,$cm$^{-3})$ and they accrete to molecular clouds with a velocity of $~10\,$km/s. We thus adopt the parameters, $\bar{n}_\mathrm{H,0}, V_0$, and $\sin\theta$, as listed in Table \ref{tab:Table1} and choose N10V10sin020 as our fiducial model. The inclination angle between the initial gas velocity and the magnetic field ($\theta$) is a key parameter to determine the physical evolution of compression layers. In order to investigate the dependence of \textrm{physical and} chemical evolution on $\theta$, we vary $\sin\,\theta$ from 0.05 (almost parallel to the initial velocity) to 1.0 (orthogonal to the initial velocity). $V_0$ is also a free parameter. In the model with higher $V_0$, the gas accumulates faster in the compression regions. We also investigate the case with a lower initial hydrogen number density. For this model, the initial velocity is set to satisfy $\bar{n}_\mathrm{H,0}V_0=10^7\,$cm$^{-2}$ s$^{-1}$, so that the growth timescale of visual extinction in the compression regions is similar to the fiducial model.

%\textrm{As we will see in Section 3, we change other parameters such as cosmic-ray ionization rate and the abundance of heavy metal and PAHs in our post-processing chemical network calculations. However, we fix these parameters in the MHD simulations as HMCR17PAHT, described in Section 3, to save computational time. This treatment can be justified because these parameters may not significantly affect the results. If we change $\zeta_\mathrm{CR}$, the gas temperature in dense and inner regions may vary, but it will remain to be a few factors \citep[e.g.][]{Clark_2019}. If we adopt $\zeta_\mathrm{CR}=1\times 10^{-16}\,\mathrm{s^{-1}}$ in equation (\ref{equ:heating}), the mean gas temperature at the center of the compression layer changes by $\sim 2\,\mathrm{K}$, the effect of which is negligible for physical structure. The heating function we adopted also depends on the existence of PAHs, but we fixed that because it also has uncertainty; e.g. it depends on the size distribution os small grain and PAHs.}

\subsubsection{Structures of the compression layer} \label{sec:struc_of_layer}

Figure \ref{fig:densityslice} shows the number density of hydrogen nuclei on $z=0$ plane at 5 Myr in the models of N10V10sin005, N10V10sin020, and N10V10sin050. The structure of the compression region becomes a quasi-steady state \textrm{in less than 1 Myr after the onset of the calculation}. Then the compression layer expands with almost constant velocity due to the continuous flow. In the models of a small inclination angle ($\sin\,\theta=0.05$), the compression layer becomes highly turbulent due to the nonlinear thin-shell instability \citep{Vishniac_1994}. The layer has a wide density distribution, and both clumpy and sparse regions appear. When the inclination is large ($\sin\,\theta=0.5$), magnetic pressure dominates, and turbulence is suppressed. With a moderate inclination angle ($\sin\,\theta=0.2$), the magnetic field suppresses the instability, but magnetic pressure is not strong enough to prevent compression. In such conditions, the compression region becomes thin and the gas density becomes high compared with the models with small or large inclinations. These features of the compression layer and their dependence on $\sin \theta$ are consistent with \citet{2019ApJ...873....6I}, who showed that the most efficient compression is achieved at the critical angle $\theta_\mathrm{cr}$.

\begin{figure*}
\includegraphics[width=1.9\columnwidth]{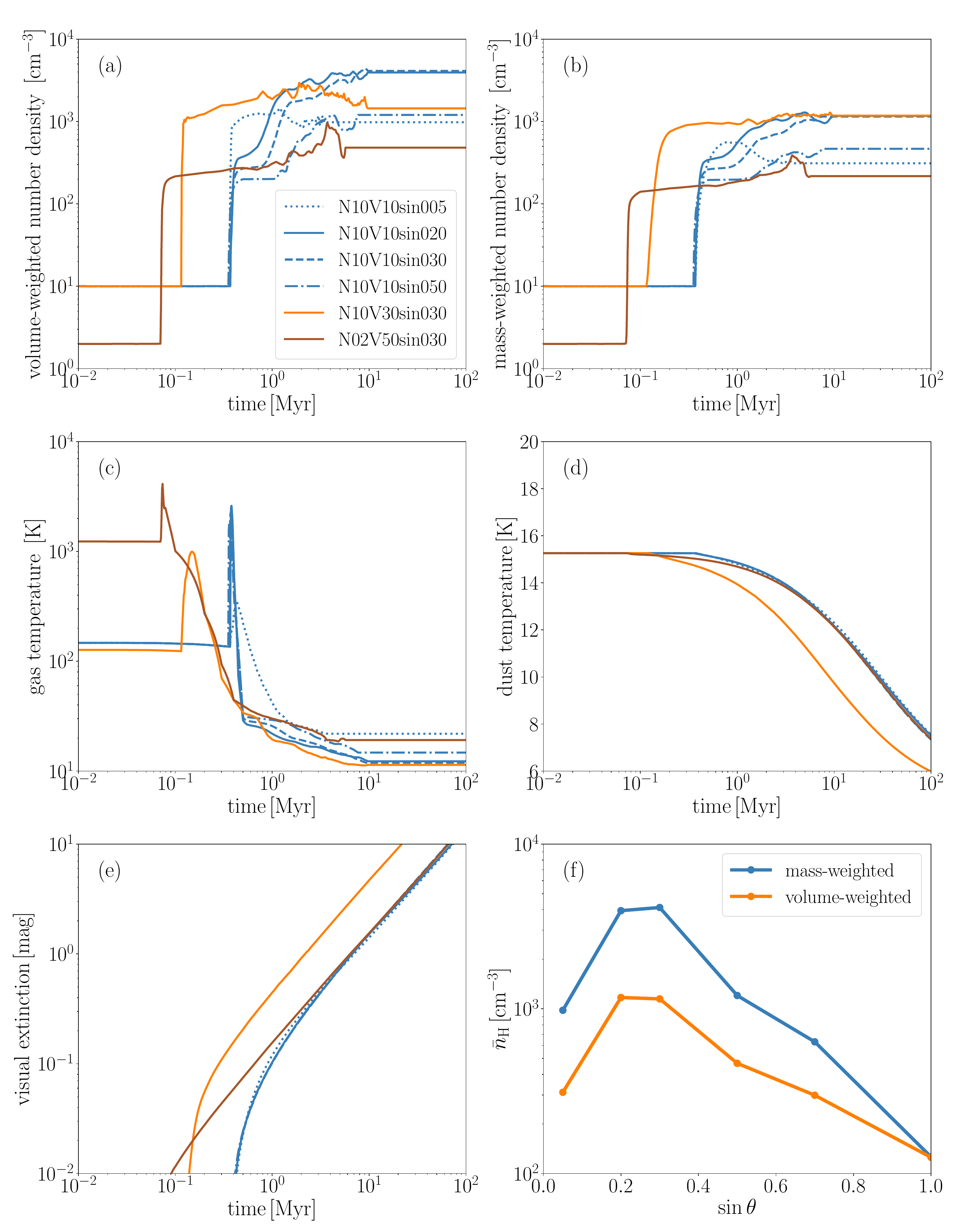}
\caption{The time evolution of (a) volume-weighted hydrogen number density, (b) mass-weighted hydrogen number density, (c) gas temperature, (d) dust temperature, and (e) visual extinction of the test particle in assorted models. \textrm{Panel (f) shows the averaged hydrogen number density of the test particle at 10 Myr in the models with $\bar{n}_\mathrm{H,0}=10\,$cm$^{-3}, V_0=10\,$km\,s$^{-1}$ as a function of $\sin\theta$. The blue and orange lines show the volume-weighted density and mass-weighted density respectively.}}
\label{fig:physicalmodel_all}
\end{figure*}

% \begin{figure}
% \includegraphics[width=\columnwidth]{figure/sin_vs_meanrho.pdf}
% \caption{The averaged density of models with $\bar{n}_\mathrm{H,0}=10\,$cm$^{-3}$ and $V_0=10\,$km\,s$^{-1}$ at 10\,Myr. The blue and orange lines show the volume-weighted density and mass-weighted density respectively.}
% \label{fig:sin_vs_meanrho}
% \end{figure}

\subsection{Analysis of the gas flow}
\label{sec:analysis_flow}

\subsubsection{Conversion of the 3D model to the 1D model}
\label{sec:analysis_flow_1dmodel}

In the present study, we investigate the detailed chemical evolution of a test particle (i.e. fluid parcel) advecting at the mean velocity field along the $x$-axis. Here, we describe how we obtain the temporal variation of the physical parameters of the test particle.

Firstly, we calculated the one-dimensional structure of the converging flow. We focus on the left side of the simulation box. We average the hydrogen number density $n_\mathrm{H}$, gas temperature $T_\mathrm{gas}$, and $x$-component of the velocity field $v_x$ along the $y$ and $z$-direction at each position $x$ and time,
\begin{equation}
    \bar{X}_\mathrm{M}(x,t)=\frac{\iint n_\mathrm{H}(x,y,z,t)X(x,y,z,t)\,\dd y\dd z}{\iint n_\mathrm{H}(x,y,z,t)\,\dd y\dd z},
\end{equation}
where $X$ represents either $n_\mathrm{H}$, or $T_\mathrm{g}$, or $v_x$. In addition to the mass-weighted mean density $(\bar{n}_\mathrm{H,M}(x,t))$, we also define the volume-weighted mean density as 
\begin{equation}
    \bar{n}_\mathrm{H,V}(x,t)=\frac{\iint n_\mathrm{H}(x,y,z,t)\,\dd y\dd z}{\iint \,\dd y\dd z}.
\end{equation}

While the volume-weighted density is appropriate for the calculation of $A_V$, the mass-weighted density would be more appropriate to investigate the chemical evolution of gas, since dense regions contain more mass, and since the chemical timescale could sensitively depend on density (see Section \ref{sec:parameter_dependence}). 
Figure \ref{fig:analysis} shows the mean values of $n_\mathrm{H}, T_\mathrm{gas}, v_x$ in model N10V10sin020. The dense compression region is formed around $x=0$ and it expands with time. The mass-weighted mean temperature gets high at the boundary of the compression region due to the shock heating, but it immediately decreases behind the shock front by radiation cooling.

Second, we calculate a trajectory of the test particle that moves with $\bar{v}_x$ as depicted with the white solid lines in Figure \ref{fig:analysis}. We put the test particle at the leftmost $x$ boundary $x=-4\,\mathrm{pc}$ at $t=0.3\,\mathrm{Myr}$, so that the test particle reaches the shock front after the post-shock structure reaches a quasi-steady state.
Then we obtain the mean density and gas temperature along the trajectory. We obtain the visual extinction by integrating the volume-weighted hydrogen number density from the shock front to the test particle. We also evaluate the dust temperature as a function of the visual extinction using equation (8) in \citet{2017A&A...604A..58H}, assuming $G_0 = 1.0$ at $A_V=0$. Since the structure of the expanding compression layer is \textrm{already} in a steady state around $10\,\mathrm{Myr}$, the physical parameters of the test particle after $10\,\mathrm{Myr}$ is obtained by extrapolation; i.e. the density and gas temperature are constant and $A_V$ increases at the rate of $\bar{n}_\mathrm{H,0}V_0$.

\subsubsection{Temporal variation of the physical parameters of a test particle}

Figure \ref{fig:physicalmodel_all} shows the temporal variation of physical parameters of the test particle in assorted models. The test particle reaches the shock front around $\sim 10^5\,\mathrm{yr}$ and experiences compression and temporal heating. As the visual extinction increases, the test particle is shielded from the external radiation field, and gas and dust temperatures decrease. 

The time evolution of physical quantities is different among models. Let us first focus on models with $\bar{n}_\mathrm{H,0}=10\,$cm$^{-3}$ and $V_0=10\,$km\,s$^{-1}$; i.e. the blue lines in Figure \ref{fig:physicalmodel_all}. The hydrogen number density at $\ga 1$ Myr, i.e. the density of the compression layer in the quasi-steady state, depends on the initial inclination angle of the magnetic field. \textrm{Panel (f) in Figure \ref{fig:physicalmodel_all}} shows the hydrogen number density of the test particle at 10\,Myr as a function of $\sin\theta$. Both the mass-weighted and volume-weighted mean densities have a peak around $\sin\theta \sim 0.2$. 
This angle is slightly smaller than the critical value expected from equation (32) in \citet{2019ApJ...873....6I}. Since we set the smaller initial density dispersion than that of \citet{2019ApJ...873....6I}, who adopted a two-phase initial condition, the compression layer in the present work becomes less turbulent, which results in a smaller critical angle. However, the basic physics is the same (see also Section \ref{sec:struc_of_layer}).

In Figure \ref{fig:physicalmodel_all}, we also compare the evolution of mean density in models with different $V_0$: N10V10sin030 and N10V30sin030. Since the growth timescale of visual extinction is proportional to \((\bar{n}_\mathrm{H,0}V_0)^{-1}\), the visual extinction increases faster in model N10V30sin030. We note that the volume-weighted mean density is comparable between the two models and the mass-weighted mean density is smaller in model N10V30sin030. This is in contrast to the 1D shock model, in which the post-shock density is higher with a higher gas velocity. It can be explained by the multi-dimensional evolution of compression layers. According to \citet{2019ApJ...873....6I}, $\sin\theta_\mathrm{cr}$ is proportional to shock velocity $V_0$ (see their equation (32)). 
The value of $\sin\theta_\mathrm{cr}$ is $\sim 0.2-0.3$ for $V_0=10$ km s$^{-1}$ \textrm{(panel (f) in Figure \ref{fig:physicalmodel_all})}, while it should be $\sim 0.6$ for $V_0=30$ km s$^{-1}$. With a fixed value of $\sin \theta =0.3$, the compression is less efficient for $V_0=30$ km s$^{-1}$. 

Finally, we compare the models of N10V10sin030 and N02V50sin030 to investigate the dependence on the initial gas density. In Figure \ref{fig:physicalmodel_all}, we can see that a smaller initial density results in a smaller post-shock density. Since we set the same initial magnetic field strength for all models, $|\vb*{B}|/\bar{n}_\mathrm{H,0}$ is larger in N02V50sin030 than in N10V10sin030. Then the magnetic pressure tends to dominate the post-shock layer and the mean density in the compression layer remains smaller in model N02V50sin030. Since we set the same value of \(\bar{n}_\mathrm{H,0}V_0\) for these models, the time evolution of visual extinction is similar.

\begin{table}
	\centering
	\caption{Initial abundances used in our chemical network calculations. HM is the same as "EA2" in \citet{2008ApJ...680..371W}.}
	\label{tab:Table2}
	\begin{tabular}{lcc} % four columns, alignment for each
		\hline
		{  } &  {High Metal (HM)}&  {Low Metal (LM)} \\
		\hline
        H       & $1.00$     & $1.00$     \\
        He      & $9.00(-2)$ & $9.00(-2)$ \\
        N       & $7.60(-5)$ & $7.60(-5)$ \\
        O       & $2.56(-4)$ & $2.56(-5)$ \\
        C$^+$   & $1.20(-4)$ & $1.20(-4)$ \\
        S$^+$   & $1.50(-5)$ & $1.50(-7)$ \\
        Si$^+$  & $1.70(-6)$ & $1.60(-8)$ \\
        Fe$^+$  & $2.00(-7)$ & $2.00(-9)$ \\
        Na$^+$  & $2.00(-7)$ & $2.00(-9)$ \\
        Mg$^+$  & $2.40(-7)$ & $2.40(-9)$ \\
        Cl$^+$  & $1.80(-7)$ & $1.80(-9)$ \\
        P$^+$   & $1.17(-7)$ & $1.17(-9)$ \\
        F$^+$   & $1.80(-8)$ & $1.80(-10)$ \\
		\hline
	\end{tabular}
\end{table}

\begin{table}
	\centering
	\caption{Chemical parameters in our chemical network calculations.}
	\label{tab:Table3}
	\begin{tabular}{lccc} % four columns, alignment for each
		\hline
		Model Name & \textrm{init. ab.} & $\zeta_\mathrm{CR}\,(\mathrm{s}^{-1})$ & PAH chemistry\\
		\hline
        HMCR16PAHT (fiducial) & HM & $1\times 10^{-16}$ & True \\ 
        LMCR16PAHT & LM & $1\times 10^{-16}$ & True \\ 
        HMCR17PAHT & HM & $1\times 10^{-17}$ & True \\ 
        HMCR15PAHT & HM & $1\times 10^{-15}$ & True \\ 
        HMCR16PAHF & HM & $1\times 10^{-16}$ & False \\ 
		\hline
	\end{tabular}
\end{table}

\section{Chemical Model}
\label{sec:chem}

We adopt a three-phase model considering gas phase, ice surface, and inert bulk mantle, using the rate equation method. We used the calculation code described in \citet{2017A&A...599A..40F}.
\textrm{The dust-to-gas mass ratio is set to be 0.01}.

The chemical reaction network and the binding energies of atoms and molecules on grain surfaces are based on KIDA\footnote{The reaction lists and the table of binding energy are taken from the default one of Nautilus code \citep{Ruaud_2016} \url{https://forge.oasu.u-bordeaux.fr/LAB/astrochem-tools/pnautilus/-/tree/master/example_simulation}}. We adopt the ratio of diffusion barrier to desorption energy as 0.4. The chemical desorption probability is assumed to be 1$\%$, following \textrm{\citet{Garrod_2007}}. The photodesorption yield is set to be $10^{-4}$ \citep{Ruaud_2016}. Since we are interested in low $A_{\rm v}$ regions, we added photodissociation and photoionization reactions referring to \citet{Garrod2013} and KIDA website, so that these reactions are considered for all neutral species \textrm{both in the gas phase and the ice surface}.
In addition, we include polycyclic aromatic hydrocarbons (PAHs). We refer to \citet{2008ApJ...680..371W} for their reactions in our fiducial calculations, the number of carbon atoms of each PAH (30 atoms), and the PAH abundance ($3.07\times 10^{-7}$ relative to total hydrogen). For simplicity, we do not consider the size distribution of PAHs. While \citet{2008ApJ...680..371W} take into account only neutral and negatively charged PAHs, we also consider positively charged PAHs. Cations in the gas phase react with neutral and negatively charged PAHs, while the anions react with neutral or positively charged PAHs to be neutralized. Positive PAHs recombine with electrons. The rate coefficients of these reactions are given in \citet{1987ApJ...320..803D}. The products of the recombination reactions are set to be the same as those of the gas-phase electron recombination. The radiative electron attachment to neutral PAHs and the photodetachment of electrons from PAHs are treated in the same way as \citet{2008ApJ...680..371W}. 

The initial chemical abundance \textrm{in the gas phase} is set to be completely atomic (Table \ref{tab:Table2}). \textrm{While we assume the Solar metallicity as the total (gas and solids) elemental abundance in the present work, there }has been a long discussion on how the elemental abundance in the gas phase varies when the interstellar gas evolves from diffuse clouds to molecular clouds. The gas-phase elemental abundance in diffuse clouds is determined by the observations of absorption lines of atoms and ions \citep[e.g.][]{1975ApJ...197...85M,1993ApJ...402L..17C,1998ApJ...493..222M}, and is called high metal abundance (HM hereinafter) in astrochemistry. It is well known that the molecular abundances in molecular clouds are better reproduced by reaction network models if the \textrm{gas-phase} abundances of Sulfur and metals (e.g. Mg, Si, and Fe) are set to be lower than those of the HM abundance by two orders of magnitudes \citep[e.g.][]{1982ApJS...48..321G}. It is called the low-metal abundance (LM).  Recent observations show that the depletion of Sulfur should start when the visual extinctions of the interstellar clouds are around several mag \citep{2023A&A...670A.114F}, which overlap with the evolutionary stage we are interested in. While our model includes the freeze-out of gas-phase species onto dust grains and their chemical reactions on grain surfaces, the actual depletion mechanisms are still unknown. We thus adopt HM as the fiducial elemental abundance in the gas phase but also consider LM (Table \ref{tab:Table2}). HM is taken from "EA2" of Table 1 in \citet{2008ApJ...680..371W}. For LM, we decrease the abundances of sulfur and heavier elements by two orders of magnitudes. \textrm{Hereafter, we call these elements, i.e. heavier than oxygen, "heavy metals".}

We set $G_0=1$ in the present work. \textrm{As described in section 2, gas temperature and dust temperature weakly depend on $G_0$, but they would not significantly affect the results of the chemical network calculation, as long as we focus on $G_0 \le 10$. The photodissociation and photoionization rate is proportional to $G_0$, and the molecular abundance as a function of $A_V$ is shifted to larger $A_V$, if $G_0$ is set to a higher value (see section 4.2.4).} 

We set $\zeta_\mathrm{CR}=1\times 10^{-16}\,\mathrm{s^{-1}}$ as the fiducial value in our chemical reaction network \citep{Indriolo_2007}. 
%We note that $\zeta_\mathrm{CR}=1\times 10^{-17}\,\mathrm{s^{-1}}$ is assumed in the heating function \textrm{used in our MHD simulations (see Section \ref{sec:basicequ})}. If we adopt $\zeta_\mathrm{CR}=1\times 10^{-16}\,\mathrm{s^{-1}}$ in equation (\ref{equ:heating}), the mean gas temperature changes by $\sim 2\,\mathrm{K}$, the effect of which is negligible for physical structure \textrm{(see also Section \ref{sec:initialparameters})}. 
The dependence of chemistry on cosmic-ray ionization rate is discussed in Section \ref{sec:chemparams}. Other input parameters for the rate equation, i.e. the total hydrogen number density, gas temperature, dust temperature, and visual extinction are given as functions of time (Figure \ref{fig:physicalmodel_all}). In order to evaluate self-shielding factors of H$_2$, C, CO, and N$_2$, we calculate the column density of each species in the post-shock gas, assuming a steady-state flow. As shown in panel (e) of Figure \ref{fig:physicalmodel_all}, visual extinction (i.e., total hydrogen column density) of our physical model monotonically increases with time. The column density of species X can then be evaluated by
\begin{equation} \label{equ:molcolumn}
    N(\mathrm{X},\,A_V)=\int_{{A_V}=0}^{A_V}\,f(X,\,A_V)N_\mathrm{H,0}\,\dd A_V,
\end{equation}
where $f(\mathrm{X},\,A_V)$ is the fractional abundance of chemical species X at each visual extinction.

Table \ref{tab:Table3} summarizes the parameters for our chemical network calculations. As we mentioned above, we adopted HMCR16PAHT, i.e. HM, $\zeta_\mathrm{CR}=1\times 10^{-16}\,\mathrm{s^{-1}}$, and include PAH chemistry in our fiducial model. We calculate four additional models to investigate the effect of \textrm{heavy metal abundance}, $\zeta_\mathrm{CR}$, and PAH (see \S\ref{sec:chemparams}).

\textrm{We note that these chemical parameters are fixed to HMCR17PAHT in our MHD simulation. In other words, the chemical parameters are varied only in the post-process chemical network calculation. The reason for this simplification is twofold. Firstly, the computational cost of our 3D MHD simulation (several $10^5$ core-hour on a supercomputer) is much more expensive than that of reaction network calculation, which is several hours with a single core on a work station. Secondly, the chemical parameters have little effect on the physical models. The choice of HM or LM is irrelevant to the MHD simulations; while it determines the abundance of heavy metals in the gas phase, they are not the main coolant. If we adopt $\zeta_\mathrm{CR}=1\times 10^{-16}\,\mathrm{s^{-1}}$ in equation (\ref{equ:heating}), the mean gas temperature at the center of the compression layer changes only by $\sim 2\,\mathrm{K}$, since the cooling function has an exponential temperature dependence. Similarly, dependence of the gas temperature and dynamics on the existence of PAHs is minor, while the heating rate may decrease by a factor of a few without PAHs \citep{Bakes1994}. In reality, these chemical parameters would vary as a function of $A_V$.} 
%According to the MHD simulations of molecular cloud formation in low-metallicity environments, reduction of the heating and cooling rates by a factor of several does not change the physical properties of the compression layers \citep{Kobayashi2023}. Our MHD simulations (with PAHs) show that the mean gas temperature is 10-30 K in the compression layer. Without PAHs, the photoelectric heating rate decreases but the gas temperature would remain \ge 10 K due to other heating mechanisms such as cosmic ray heating. 

%If we change $\zeta_\mathrm{CR}$, the gas temperature in dense and inner regions may vary, but it will remain to be a few factors \citep[e.g.][]{Clark_2019}. If we adopt $\zeta_\mathrm{CR}=1\times 10^{-16}\,\mathrm{s^{-1}}$ in equation (\ref{equ:heating}), the mean gas temperature at the center of the compression layer changes by $\sim 2\,\mathrm{K}$, the effect of which is negligible for physical structure. The heating function we adopted also depends on the existence of PAHs, but we fixed that because it also has uncertainty; e.g. it depends on the size distribution os small grain and PAHs.}

\section{Result} \label{sec:result}

\subsection{Fiducial model}
\label{sec:fiducial}

\begin{figure*}
\includegraphics[width=1.9\columnwidth]{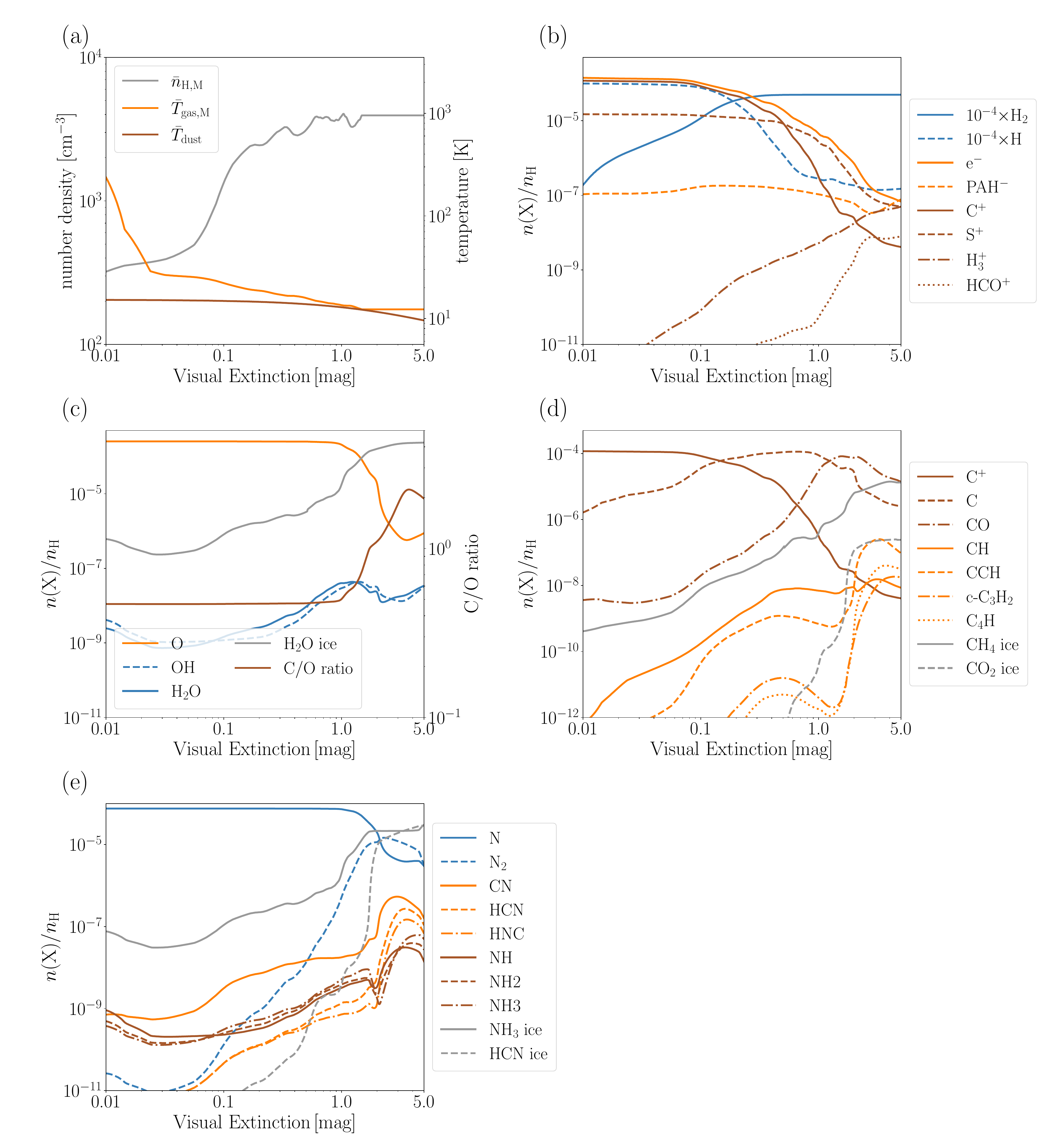}
\caption{(a) The total hydrogen number density, gas temperature, and dust temperature of model N10V10sin020, as a function of visual extinction. (b)-(e) The abundances relative to hydrogen nuclei. Each panel shows hydrogen, electron, PAH$^-$, and major cations, Oxygen-bearing species, Carbon-bearing species, and Nitrogen-bearing species. The abundances of atomic and molecular hydrogen are multiplied by $1\times 10^{-4}$. In panel (c), the C/O ratio in the gas phase is also plotted.}
\label{fig:N10V10sin020_mainspecies}
\end{figure*}

\begin{figure}
\includegraphics[width=\columnwidth]{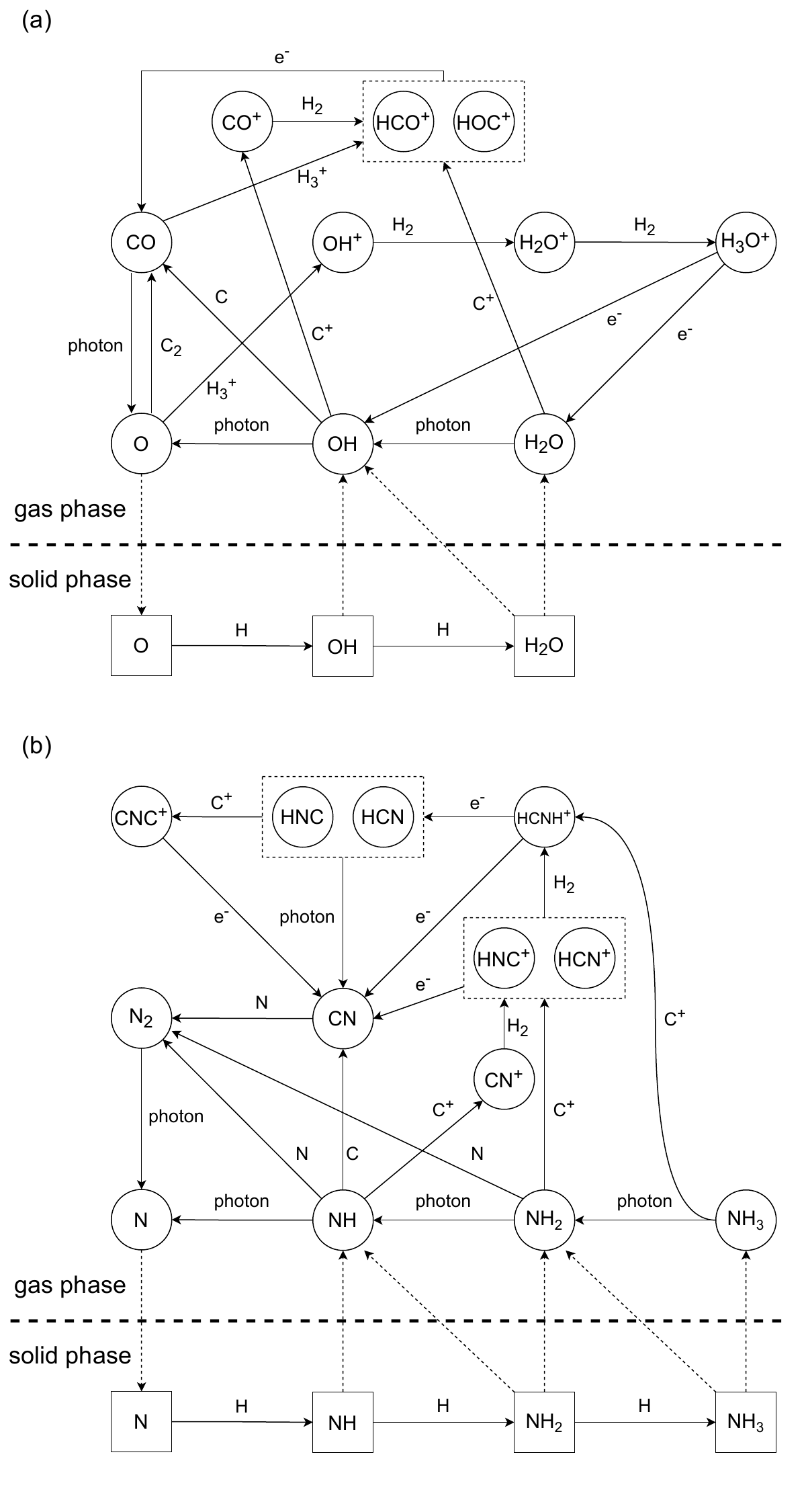}
\caption{The major chemical reactions for \textrm{O-bearing (a) and} N-bearing \textrm{(b)} species at $A_V \la 1$ mag. Circles and squares indicate chemical species in the gas phase and solid phase, respectively. Dashed lines show the adsorption and desorption.}
\label{fig:Nroute}
\end{figure}

In this section, we show the results of chemical network calculation with the fiducial parameter sets (model N10V10sin020 and HMCR16PAHT). Figure \ref{fig:N10V10sin020_mainspecies} shows the evolution of the physical parameters and fractional abundances of selected chemical species with respect to hydrogen nuclei. Since $A_V$ increases with time (see panel (e) of Figure \ref{fig:physicalmodel_all}), we show them as functions of $A_V$. 
Panel (a) presents the total hydrogen number density, gas temperature, and dust temperature. As the gas density increases, radiative cooling becomes efficient, and the gas temperature reaches around 20 K. The present work focuses on the early stage of molecular cloud formation and does not include self-gravity in the MHD simulations. Although we conducted the chemical network calculations up to $A_V = 10\,\mathrm{mag}$, where self-gravity becomes important, we mainly discuss results at $A_V \la 3\,\mathrm{mag}$.

Figure \ref{fig:N10V10sin020_mainspecies} (b) shows the abundances of atomic and molecular hydrogen, electrons, and major ions. \textrm{The HI-to-H$_2$ transition occurs} at $A_{\rm V} \ga 0.1\,\mathrm{mag}$, which is consistent with \citet{2004ApJ...612..921B}, \citet{2010A&A...515A..66H}, and \citet{2015A&A...584A.124F}.
The main charge carrier also changes with $A_{\rm V}$. The initially dominant charge carriers, C$^+$ and S$^+$, start to decrease at $A_{\rm V} \sim$ 0.1 mag and 1 mag, respectively, due to recombination with PAHs. In this range of $A_{\rm V}$, $30-50\,\%$ of PAHs is negatively charged. 

The cosmic-ray ionization starts to dominate over photoionization around $A_V\sim 1$ mag. The ionization degree also decreases accordingly. At $A_V \sim 3$ mag, H$_3^+$ starts to dominate the positive charge, and the electron abundance decreases to $10^{-7}$. At $A_V>3$ mag, the abundance of PAH$^-$ is comparable to that of electrons. 
Since the effective formation pathway for HCO$^+$ is the proton transfer reaction between CO and H$_3^+$, the fractional abundance of HCO$^+$ also increases following CO and H$_3^+$.

Figure \ref{fig:N10V10sin020_mainspecies} (c) shows the evolution of O-bearing species\textrm{, and Figure \ref{fig:Nroute} (a) displays the major formation and destruction routes of O-bearing species at $A_V\lesssim 1$.} 
Oxygen is initially in the atomic form. At $A_V\sim$ a few mag, water ice becomes the dominant oxygen reservoir, and its abundance saturates to $\sim 10^{-4}$ around 3 mag. This trend is consistent with \citet{2010A&A...515A..66H}. %ur results. 
OH and H$_2$O in the gas phase are basically formed by ion-neutral reactions initiated by O $+$ H$_3^+$. At $A_V< 1$ mag, however, chemical desorption of H$_2$O ice also contributes to the abundance of H$_2$O in the gas phase. OH is also formed by the photodissociation of gaseous H$_2$O. As a result, the abundances of gaseous H$_2$O and OH show a similar temporal variation to that of water ice.

Panels (d) and (e) in Figure \ref{fig:N10V10sin020_mainspecies} show the abundances of C and N-bearing molecules, respectively. Here we summarize the trend of C-bearing molecules without oxygen (e.g. CCH and CN) in our models. The formation of those molecules depends on the major carbon budgets (i.e. C$^+$, C, and CO). When neutral atomic carbon is abundant, the formation of the C-bearing molecules proceeds as well. At 0.5 mag $< A_V < 1$ mag, carbon is consumed by CO formation, and the abundance of carbon chains decreases. This threshold of visual extinction for CO formation is consistent with \citet{2004ApJ...612..921B}. However, as oxygen depletes from the gas phase to form water ice, the elemental C/O ratio is enhanced in the gas phase, and the carbon chains, CN, and HCN increase \citep{Terzieva_1998, Aikawa_2003,Ruaud_2018}. At $A_V>3\,\mathrm{mag}$, CO reacts with He$^+$ to supply C$^+$ to the chemical network, but  C$^+$ is neutralized by the reaction with PAH$^-$ and then adsorbed onto grains. Due to hydrogenation reactions, ice species of hydrocarbons such as CH$_4$ become the major carbon carriers.

Figure \ref{fig:N10V10sin020_mainspecies} (e) shows the evolution of N-bearing species.
The dominant reservoir of nitrogen is N atoms in the gas phase at $A_V\le 1-2$ mag. At higher visual extinction, NH$_3$ ice and HCN ice dominate. Note that the relative abundances of such ice species sensitively depend on the diffusion barrier of hydrogen atoms on the grain surface, which is not well constrained. If the diffusion barrier is set to be lower than the fiducial value ($230$ K), which is close to the experimental value \textrm{\citep{Hama_2012}}, the hydrogenation reactions become more efficient. This results in a higher abundance of NH$_3$ ice, while HCN ice, which is mainly formed by CN + H$_2$ on grain surface, becomes less abundant. In addition, \citet{Rimola_2018} suggest that adsorbed CN can react with water ice to form formamide (NH$_2$CHO). If this reaction is considered, the abundance of HCN ice may decrease.

It is noteworthy that grain surface reactions contribute to the formation of simple N-bearing molecules such as N$_2$ and CN at low visual extinction, while they are solely formed by neutral-neutral reactions such as CH + N $\rightarrow$ CN + H in the gas phase in classical pseudo-time dependent models \citep[e.g.][]{1980ApJ...239..151P,1980ApJS...43....1P,1982ApJS...48..321G}. 
\textrm{Panel (b)} in Figure \ref{fig:Nroute} shows the major formation paths of CN and N$_2$ at $A_V\la 1\,\mathrm{mag}$. Atomic nitrogen is adsorbed onto grain surfaces and hydrogenated to form NH$_3$ ice. Chemical desorption of NH$_3$ and its precursor 
 enhances the gas-phase abundance of NH and NH$_2$, which react with N atom to form N$_2$. They also react with C$^+$ and atomic C to form HCN and HNC. The importance of NH$_3$ ice formation for CN and N$_2$ is also discussed by e.g. \citet{Wagenblast_1996}, \citet{2010A&A...520A..20G} and \citet{2018MNRAS.476.4994F}.

At $A_V \ga 1\,\mathrm{mag}$, CN and HCN are formed via reactions between N atom and hydrocarbons. As the elemental C/O ratio increases in the gas phase, CN and HCN increase significantly together with hydrocarbons. 

The temporal variation of hydrocarbons, CN, and HCN described above is different from that of the classical pseudo-time-dependent models which assume $A_V\ga 3$ mag. In The pseudo-time-dependent models, their abundances reach the maximum value when neutral atomic carbon is the dominant carbon reservoir, and they decrease as CO dominates.
The reason for the different behavior of our model is the temporal variation of the visual extinction (Figure \ref{fig:physicalmodel_all}). When neutral atomic carbon is abundantly formed, the visual extinction is still low ($\la 1$ mag), and the photodissociation of the C-bearing molecules competes with their formation.

Molecular abundances in our model are similar to those of steady-state PDR models at low $A_{\rm V}$ where the photodissociation keeps the chemical timescale shorter than the dynamical timescale. 
Differences become apparent at higher $A_{\rm V}$. We also note that grains-surface reactions which play important roles in our model are not considered in many PDR models. For example, our model shows higher abundances of CN and HCN around $A_V< 1\,\mathrm{mag}$, possibly due to reactions in Figure \ref{fig:Nroute} (b), than the static PDR models \citep[e.g.][]{Boger_2005}.
Hydrocarbons increase as gaseous C/O ratio is enhanced due to H$_2$O ice formation in our model, while their abundances reach the maximum right before the conversion of atomic carbon to CO in the PDR models \citep[e.g.][]{Teyssier_2004,Pety_2005}. 

\subsection{parameter dependence}
\label{sec:parameter_dependence}

We calculate chemistry in 8 shock models in total (Table \ref{tab:Table1}) to investigate the dependence of molecular evolution on shock parameters, i.e. initial inclination angle between the conversing flow and magnetic field, velocity, and initial gas density. We also investigate the dependence of molecular evolution on chemical parameters (Table \ref{tab:Table3}).

\begin{figure*}
\includegraphics[width=1.9\columnwidth]{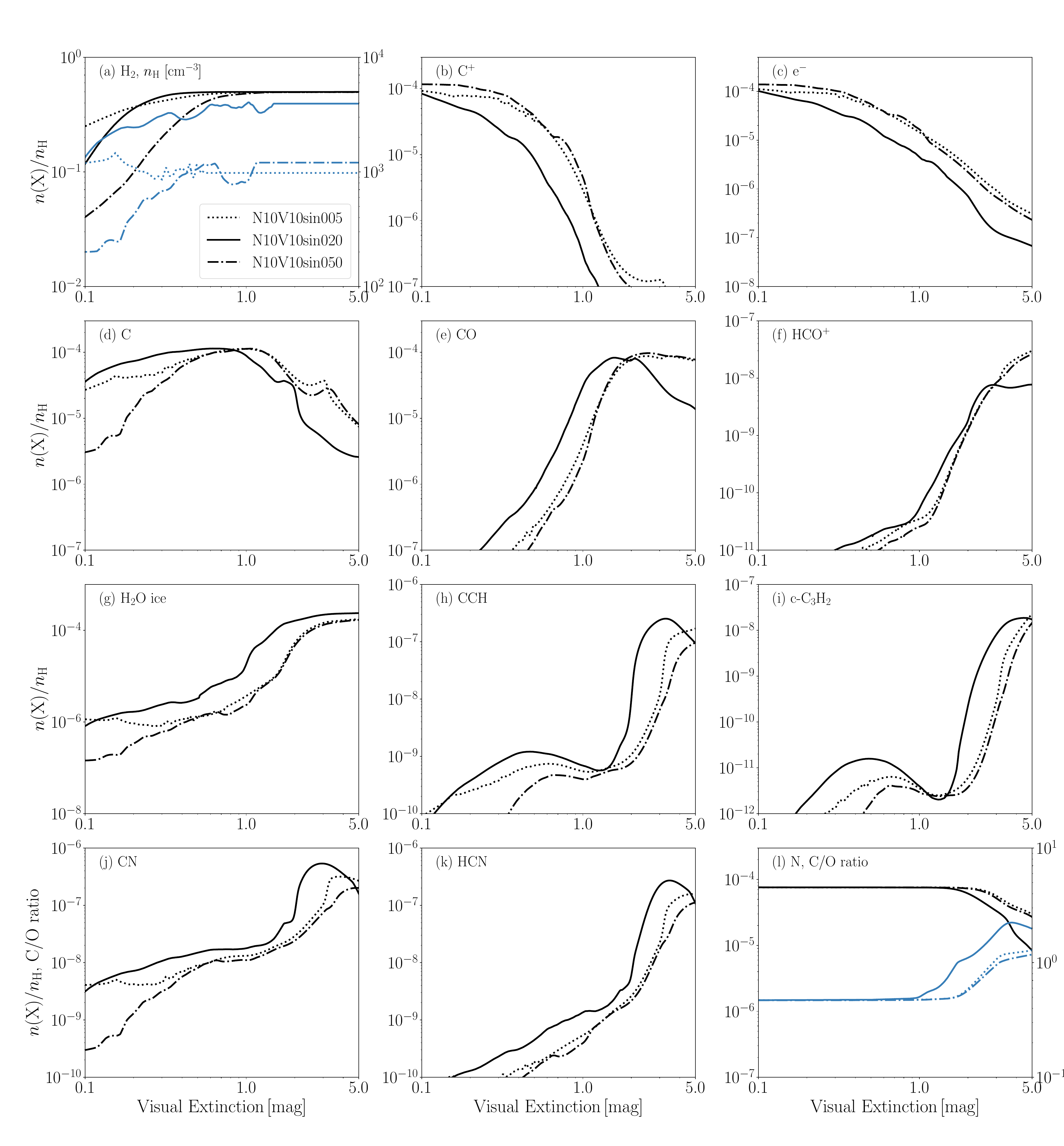}
\caption{The fractional abundance of chemical species and the C/O ratio in the gas phase for the models with different initial inclination of the magnetic field. \textrm{The blue lines in panel (a) show the gas density ($n_{\rm H}$) with the value written on the second y-axis (on the right) in the unit of cm$^{-3}$. The blue lines in panel (l) with the second y-axis show the C/O ratio in the gas phase.}}
\label{fig:bfield_spec}
\end{figure*}

% \begin{figure}
% \includegraphics[width=\columnwidth]{figure/bfield_CN_v3.pdf}
% \caption{\textrm{Abundances of N atom, CN, and C-bearing molecules relevant to CN formation as functions of visual extinction in models N10V10sin020 (solid line) and N10V10sin050 (dashed line).}}
% \label{fig:bfield_hydrocarbon_CN}
% \end{figure}

\begin{figure*}
\includegraphics[width=1.9\columnwidth]{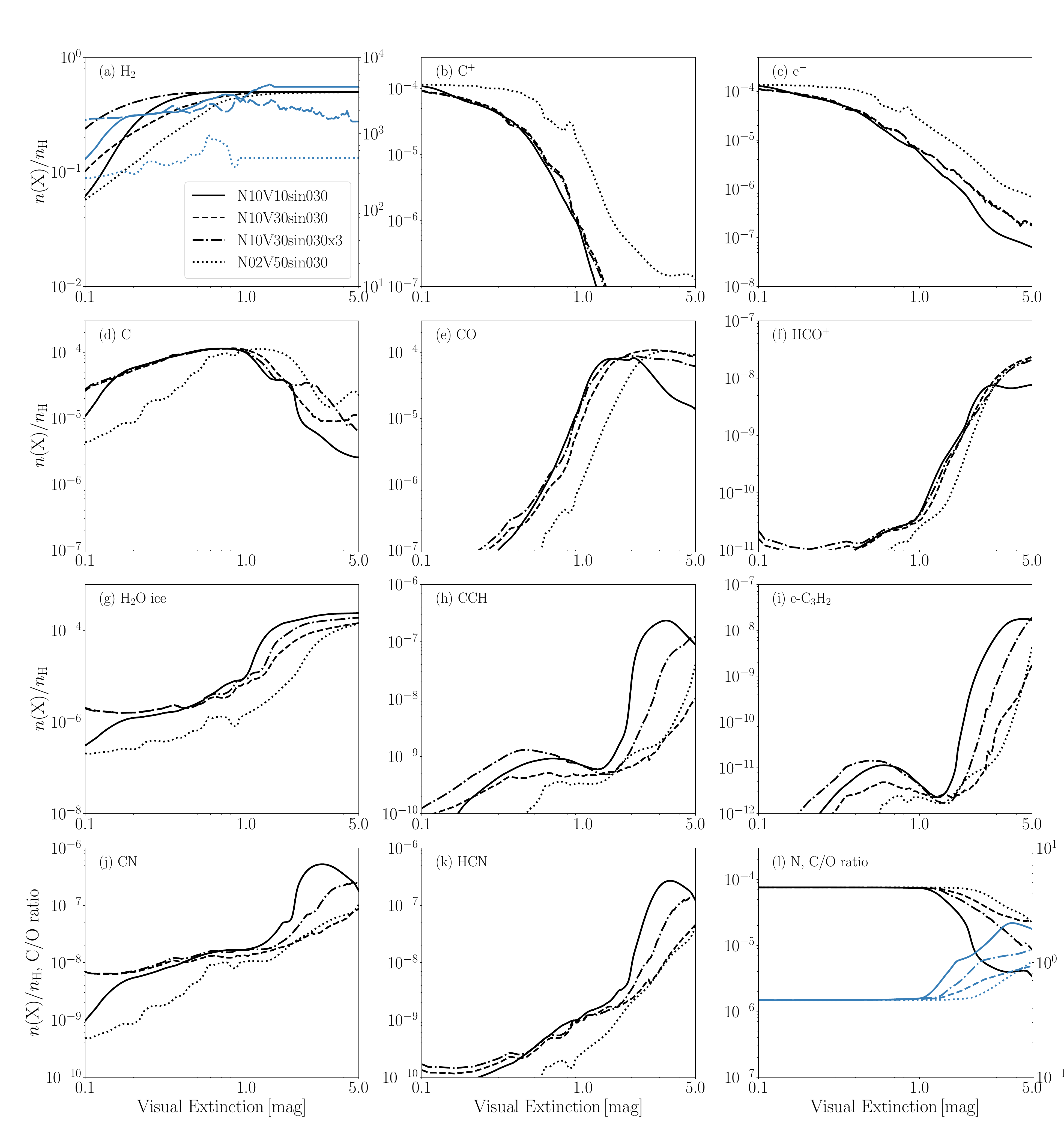}
\caption{Similar to Figure \ref{fig:bfield_spec}, but for the models with different shock velocity and initial density.}
\label{fig:velocity_density_spec}
\end{figure*}

\begin{figure*}
\includegraphics[width=1.9\columnwidth]{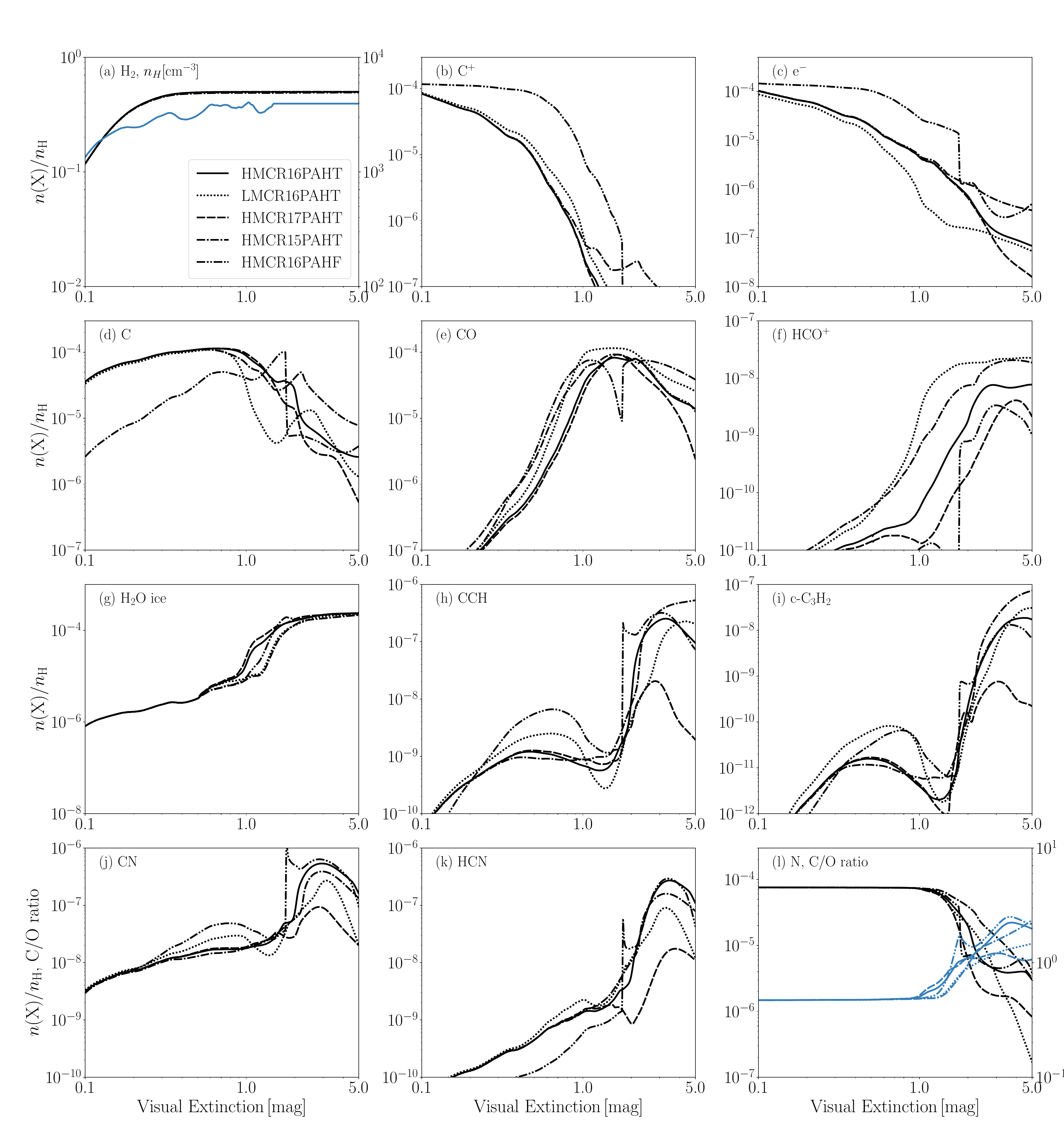}
\caption{Similar to Figure \ref{fig:bfield_spec}, but for model N10V10sin020 with various chemical parameters.}
\label{fig:chemical_param_spec}
\end{figure*}

\subsubsection{inclination of magnetic field}
\label{sec:bfield_dep}

Figure \ref{fig:bfield_spec} shows the abundances of selected species in the three models, varying the initial inclination angle of the magnetic field. The main difference between these models is the post-shock density, which is 3-4 times higher in model N10V10sin020 than in the other two models\textrm{, as shown in panel (a) with blue lines (see also Figure \ref{fig:physicalmodel_all} (f)). The rise of H$_2$ abundance correlates with that of gas density, since the H$_2$ formation rate is proportional to gas density.} 

Two-body reactions compete with photodissociation and photoionization (one-body reactions) more efficiently at higher densities. Thus the decline of C$^+$ due to neutralization by PAH$^-$ and the formation of CO occurs at lower $A_V$ in the model with $\sin\theta=0.2$, for example. \textrm{The electron abundance follows the variation of C$^+$ abundance, while the dominant positive charge carrier is switched from C$^+$ to S$^+$ at $A_V\sim 1$ mag.}
CO is formed via neutral-neutral reactions such as C + OH. The major formation path of OH is the series of gas-phase reactions initiated by H$_3^+$ + O, which are also more effective at higher densities at $A_V\la 1$ mag. \textrm{HCO$^+$ roughly follows the variation of CO but shows less dependence on post-shock density, because HCO$^+$ formation starts from the reaction C$^+$ + OH $\rightarrow$ CO$^+$ + H and it is destructed by dissociative recombination with electrons at $A_V\le 1$ mag.}

While the dependence of two-body reaction rates on density is positive in general, it varies among reaction type, e.g. adsorption onto grains and ion-molecule reactions.
Decline of CO abundance starts earlier in model N10V10sin020 than in the other two models since the adsorption rate of CO onto grain surfaces is proportional to density. We note, however, that CO depletion in model N10V10sin020 is more severe than shifting the evolution to lower $A_{\rm V}$.
In addition to the adsorption onto gains, gaseous CO is destructed by He$^+$, i,e, CO + He$^+$ $\rightarrow$ C$^+$ + O + He. Formation of the ices of H$_2$O and CH$_4$, via recombination of PAH$^-$ and adsorption onto grains of C and O, competes with CO reformation.

The post-shock density, and thus the inclination angle of the magnetic field are also important for carbon-chain chemistry. 
$\ce{CCH}$ and $\ce{c-C3H2}$ rapidly increases around $A_V\simeq 1-3\,\mathrm{mag}$ (the second bottom row of Figure \ref{fig:bfield_spec}). The visual extinction of such rapid carbon-chain increase is lower in model N10V10Bsin020 than in the other two models.
There are mainly two reasons. Firstly, abundant carbon is needed for the efficient formation of carbon chains. Carbon chains consist of more than one carbon and extend via chemical reactions with neutral and ionized carbons \citep{2013ChRv..113.8981S}. The total number density of $\ce{C+}$ and $\ce{C}$ at $A_V\le 1\,\mathrm{mag}$ is larger in the model of higher post-shock density. 
Secondly, H$_2$O ice formation, which enhances the elemental C/O ratio in the gas phase, is more efficient at higher density, since it is initiated by the \textrm{adsorption} of atomic oxygen and hydrogen. 
\textrm{Panel (g)-(i) and (l) in Figure \ref{fig:bfield_spec}} clearly demonstrate this trend; $\ce{H2O}$ ice gradually forms around $A_V\simeq 1-2\,\mathrm{mag}$, and \textrm{the C/O ratio in the gas phase and the abundances of CCH and c-C$_3$H$_2$ accordingly increase}.
Efficient formation of CH$_4$ ice, however, results in slight decline of carbon chains in the gas phase in model N10V10sin020 at $A_{\rm V}\sim 3$ mag.

The bottom left and middle panels in Figure \ref{fig:bfield_spec} show the abundances of CN and HCN, respectively. The dependence of their abundances on the initial magnetic field angles is qualitatively similar to that of carbon chains, but slightly less significant. At $A_V\la 1\,\mathrm{mag}$. the formation of CN and HCN is initiated mainly by NHn + C$^+$.
While the grain-surface formation of NH and NH$_2$ (\S 4.1) is more efficient, C$^+$ is less abundant in the model with higher post-shock density.
At $A_V\ga 1\,\mathrm{mag}$, CN and HCN are formed by the reaction of N atom with hydrocarbons. In the model with higher post-shock density, carbon chains are more abundant, but N atom depletes more efficiently onto grains.

\subsubsection{shock velocity}
\label{sec:parameter_velocity}

\textrm{The solid and dashed lines} in Figure \ref{fig:velocity_density_spec} show the abundance of selected molecules in models with the same magnetic field angle ($\sin \theta=0.3$) but with different shock velocities, $V=10$ and 30 km s$^{-1}$. It should be noted that the inclination angle of the magnetic field for efficient compression varies with the gas density and velocity. While the angle is $\sin\theta = 0.2$ for our fiducial model ($n=10\,\mathrm{cm^{-3}}, V=10\,\mathrm{cm/s}$), it is $\sin\theta = 0.3$ for the case of higher velocity or lower initial density (see \S \ref{sec:parameter_density}).
Among the two models in Figure \ref{fig:velocity_density_spec}, the lower shock velocity results in higher post-shock density, which accelerates the formation of carbon-chain molecules (see the previous section). On the other hand, CO formation occurs at almost the same $A_V$, which indicates that $A_V$ (i.e.extinction by dust) and self-shielding regulate the formation of CO.

The shock velocity also determines how much gas is injected into the compression layers; i.e. the growth rate of visual extinction is proportional to the shock velocity. In order to investigate the effect of accumulation timescale, we calculate molecular evolution in another model "N10V30sin030x3", in which the physical structure (i.e. density and temperature as a function of $A_{\rm v}$) is the same as that of N10V30sin030 but the velocity of gas (and thus the test particle) is artificially slowed down by a factor of 3. We can see the effect of the accumulation timescale by comparing the \textrm{dot-dashed} line (N10V30sin030x3) and \textrm{dashed} line (N10V30sin030) in Figure \ref{fig:velocity_density_spec}. 
For example, the abundance of \textrm{H$_2$ and} water ice in N10V30sin030x3 is larger than that of N10V30sin030. The post-shock gas density in the model is $\sim 10^3$ cm$^{-3}$, and the timescale of water ice formation ($\sim$ adsorption timescale of O atom onto grains) is $\sim$ 3 Myrs, which is slightly larger than the timescale for the test particle to reach $A_{\rm V}\sim 1$ mag ($\sim$ 2 Myr) in the model of N10V30sin030. In the model with lower velocity, more water can be formed before the test particle reaches the same $A_{\rm v}$, while other reactions could compete with water ice formation. 
Accordingly, the C/O ratio in the gas phase and the abundance of carbon chains \textrm{and CN-bearing species} increases at lower $A_{\rm v}$ in model N10V30sin030x3. \textrm{On the other hand, the abundances of C$^+$, C, and CO are less dependent on the accumulation timescale, which indicates that the reactions directly related to those species proceed faster than the gas accumulation in N10V30sin030.}

\subsubsection{initial density}
\label{sec:parameter_density}

Figure \ref{fig:velocity_density_spec} also shows the molecular evolution in models N10V10sin030 and N02V50sin030, which have similar accumulation timescale but the initial gas density is different by a factor of 5. Figure \ref{fig:physicalmodel_all} shows that the post-shock density is also lower by a similar factor in model N02V50sin030, which results in slower molecular evolution. Although \textrm{the HI-to-H$_2$ and C-to-CO transitions occur} at $A_V<3$ mag in N2V50sin030, minor molecules such as hydrocarbons are less abundant than those in N10V10sin020. At $A_V\sim 3\,\mathrm{mag}$, for example, the abundances of $\ce{CCH}$ and $\ce{c-C3H2}$ are higher in model N10V10sin030 by almost 3 orders of magnitudes. 
N-bearing molecules (i.e. CN, HCN and HNC) are also more abundant at $A_V\sim 3$ mag in model N10V10sin030, while the difference between the two models is slightly less significant than carbon chains.

\subsubsection{chemical parameters} \label{sec:chemparams}

So far, we adopted HMCR16PAHT for the chemical parameter set and investigated the dependence of molecular evolution on shock parameters. In this section, we investigate the effect of the abundance of heavy elements, cosmic ray ionization rate, and the existence of PAHs using our fiducial shock model N10V10sin020.

Figure \ref{fig:chemical_param_spec} shows the temporal variation of the abundances of selected species. 
\textrm{Solid and dotted} lines show the molecular abundances with the high metal abundance (HMCR16PAHT) and low metal abundance (LMCR16PAHT), respectively.
Compared with LM, HM sets higher abundance of heavy elements such as S and Si, which provide electrons due to photoionization or cosmic ray ionization.
Free electrons recombine with HCO$^+$, so that HCO$^+$ abundance is lower in HM than in LM. The effect of \textrm{heavy metal abundance} is mitigated by PAHs, which neutralize atomic ions \citep{2008ApJ...680..371W}.
Yet, S$^+$ is one of the major ions in model HMCR16PAHT even at $A_V\ga 1$ mag (Figure \ref{fig:N10V10sin020_mainspecies}b), %Free electrons provided by S atoms recombine with HCO$^+$, so that 
and HCO$^+$ abundance is lower in HMCR16PAHT than in LMCR16PAHT by about one order of magnitude. The difference between HM and LM becomes more significant without PAHs.

\textrm{Solid, dashed, and dot-dashed} lines in Figure \ref{fig:chemical_param_spec} depict the molecular evolution with the cosmic ray ionization rate $\zeta_\mathrm{CR}$ of $10^{-16}, 10^{-17}$ and $10^{-15}$ s$^{-1}$, respectively. At $A_V\la 1\,\mathrm{mag}$, there is little dependence on cosmic ray ionization rates, because photoionization by the interstellar UV dominates over cosmic-ray ionization. The difference between the three models becomes apparent at $A_V\ga 2\,\mathrm{mag}$. For example, molecular ions such as H$_3^+$ and HCO$^+$ become major ions and they are more abundant in the models with higher $\zeta_\mathrm{CR}$.

In the model with a low cosmic-ray ionization rate (\textrm{dashed line}), the abundances of carbon-chain molecules are lower than those of the fiducial model (\textrm{solid line}). At $A_{\rm v}\ga 2$ mag, carbon chains are formed by reactions between hydrocarbons and C$^+$, which is provided by CO + He$^+$. Since the ionization potential of He is higher than that of H, He$^+$ abundance scales with the cosmic-ray ionization rate. Interestingly, the abundance of water ice and the C/O ratio in the gas phase are almost the same in the three models. It indicates that even if the C/O is larger than unity, carbon-chain molecules cannot be more abundant than $\sim 10^{-8}$ relative to hydrogen nuclei if the cosmic ray ionization is lower than $1\times 10^{-17}\,\mathrm{s^{-1}}$. Carbon chains are slightly more abundant in the model with higher cosmic-ray ionization (dot-dashed blue line) than in the fiducial model. 
A similar effect can be seen for CN, HCN, and HNC.

The observations of H$_3^+$ show that the cosmic-ray ionization rate in diffuse clouds is $\sim 1\times 10^{-16}\,$s$^{-1}$ or even higher \citep[e.g.][]{2012ApJ...745...91I}.
%The model with a low ionization rate (HMCR17PAHT) is thus not appropriate for diffuse clouds. 
While we adopt a constant cosmic-ray ionization rate, \citet{2015ApJ...812..135I} and \citet{2018A&A...614A.111P} derived $\zeta_\mathrm{CR}$ as a function of gas column density.
According to their model $\mathscr{H}$, the $\zeta_\mathrm{CR}$ at $A_V\sim$ a few mag is still between $10^{-16}-10^{-15}$ s$^{-1}$. This indicates that the molecular abundances in diffuse clouds should be between those of HMCR16PAHT and HMCR15PAHT.

%\textrm{We note that the abundance of CO of HMCR15PAHT is larger than that of HMCR16PAHT. This trend is different from the case of dense cloud chemistry; higher $\mathscr{H}$ decreases the CO abundance \citep[e.g.][]{Bialy_2015}. There are two reasons for this trend at $A_V < 1$ mag. One is that higher $\mathscr{H}$ increases the abundance of H$_3^+$ and thus CO$^+$ (see panel (a) in Figure \ref{fig:Nroute}). The other is that photodissociation of CO works at such a low $A_V$, and the destruction by He$^+$, which is effective in higher $\mathscr{H}$ condition, does not proceed so much. A similar trend can be found in LMCR16PAHT; LM results in a smaller ionization degree, which inactivates the destructive recombination of H$_3^+$ and helps CO formation at $A_V\le 1$ mag. At $A_V > 1$ mag, the photodissociation of ice by CR-induced UV is effective and the depletion of carbon from the gas phase is suppressed, which keeps CO abundance in the gas phase.}

\textrm{We note that CO abundance in model HMCR15PAHT is higher than that in HMCR16PAHT. This is in contrast to the models of dense clouds, which predict lower CO abundance with higher $\zeta_\mathrm{CR}$ \citep[e.g.][]{Bialy_2015,Bisbas_2015,Bisbas_2017}. The reason for this difference is two fold. Firstly, at low $A_V$ ($< 1$ mag), higher $\zeta_\mathrm{CR}$ enhances H$_3^+$ abundance, and thus the formation of CO via OH and CO$^+$ (Figure \ref{fig:Nroute} (a)). Secondly, our model takes into account more detailed processes of gas-dust interaction than some of the previous work. While higher $\zeta_\mathrm{CR}$ enhances destruction of CO by He$^+$, it also enhances the photodissociation and desorption of icy molecules induced by cosmic rays, which maintains more carbon and oxygen in the gas phase.}

Fig \ref{fig:chemical_param_spec} \textrm{also shows} the molecular evolution in the model without PAHs (HMCR16PAHF). In the absence of efficient neutralization by \textrm{PAH$^-$}, C$^+$ is abundant up to $A_V\sim 1$ mag.
We note that the electron abundance suddenly decreases at $A_V\la 2$ mag. It is not a numerical artifact but due to the change of the major reactants of electrons from atomic ions such as S$^+$ to carbon-chain cations and the difference of rate coefficients between these reactions (Appendix \ref{app:bistability}).
Such sudden decline of the electron abundance does not appear in models with PAHs (e.g. HMCR16PAHT), in which the main formation and destruction paths of \textrm{electrons} are
\begin{equation}
    \ce{PAH-} + \ce{photon} \rightarrow \ce{PAH} + \ce{e-},
\end{equation}
and
\begin{equation}
    \ce{PAH} + \ce{e-} \rightarrow \ce{PAH-} + \ce{photon}.
\end{equation}
Especially, the latter reaction always dominates over the recombination with gaseous ions. This prevents the sudden change of ionization degree.
A more detailed discussion on HMCR16PAHF is in Appendix \ref{app:bistability}.

\textrm{In addition to the parameters discussed above, we also investigated the effect of $G_0$. We conducted chemical reaction network calculations with the fiducial physical and chemical parameters varying $G_0$ from unity to 10. The photodissociation and photodesorption rate linearly increases with $G_0$. The effect of $G_0$ on molecular abundances is rather simple; in the case of higher $G_0$, the layers of  C$^+$-to-C transition, water ice formation, and C/O ratio enhancement are shifted to higher $A_V$. The molecular abundances as a function of $A_V$ are also shifted accordingly.}

\section{Discussion} \label{sec:discussion}
\subsection{comparison with observations}

We compare molecular column densities obtained from our models with observations of interstellar clouds. Since we focus on the early stage of molecular cloud formation, i.e. postshock gas with $A_V\la $ several $\,\mathrm{mag}$ and $n_\mathrm{H}<10^4\,\mathrm{cm^{-3}}$, we refer to the observations of molecular absorption lines detected in diffuse and translucent clouds. 

The absorption lines give us the total column density through the interstellar cloud. We have calculated the molecular abundances along the trajectory of a test particle, assuming that the flow is almost in a steady state (Figure \ref{fig:analysis}). In order to compare our results with the observations, we derived the column density of species X from our models as follows:
\begin{equation} \label{equ:molcolumn_ete}
    N^{\rm obs}(\mathrm{X},\,A_V^\mathrm{obs})=2\int_{A_V=0}^{A_V}\,f(X,\,A_V)N_\mathrm{H,0}\,\dd A_V,
\end{equation}
where $A_V^{\rm obs}=2 A_{\rm v}$ \citep{2010A&A...515A..66H}. 
%the visual extinction measured from one edge to edge for each cloud.
We note that the column density is multiplied by a constant factor if we are observing the compression layer obliquely or if there are multiple compression layers in the line of sight. It does not significantly change the discussion below, since the molecular abundances and thus the molecular column densities increase non-linearly with $A_{\rm V}$ in a compression layer.
\textrm{Relation between $A_V^\mathrm{obs}$ and the effective visual extinction in more general 3D morphology is discussed by e.g. \citet{Clark_2014} and \citet{Bisbas_2023}.}

% \begin{figure}
% \includegraphics[width=\columnwidth]{figure/vsobs_CCH_C3H2_ete.pdf}
% \caption{Comparison of the column densities of HCO$^+$ and carbon-chain molecules between observations (circles) and our models (lines). Dotted, solid, and dashed blue lines depict the model of  N10V10sin005, N10V10sin020, and N10V10sin050, respectively. The orange solid line is N10V30sin030, and the brown solid line is N02V50sin030. The black-filled squares and triangles represent the molecular column density of each model at $A_V^{\rm obs}=1$ and $5\,\mathrm{mag}$, respectively. The filled circles indicate the observational values of \citet{1996A&A...307..237L} and \citet{2000A&A...358.1069L} (purple), \citet{2010A&A...520A..20G} and \citet{2011A&A...525A.116G} (gray), \citet{2011A&A...535A.103M} (yellow), \citet{2016PASJ...68....6A} (orange), \citet{2018A&A...610A..43R} (red), and \citet{2023A&A...670A.111K} (sky blue).}
% \label{fig:vsobs_CChain_phys}
% \end{figure}

% \begin{figure}
% \includegraphics[width=\columnwidth]{figure/vsobs_CCH_C3H2_chemparam_ete.pdf}
% \caption{Similar to Figure \ref{fig:vsobs_CChain_phys}, but the lines show the results of model N10V10sin020 with various chemistry parameters: HMCR16PAHT (solid blue line), HMCR15PAHT (dash-dot blue line), HMCR17PAHT (dashed blue line), LMCR16PAHT (solid orange line), and HMCR16PAHF (solid brown line).}
% \label{fig:vsobs_CChain_chem}
% \end{figure}

\begin{figure*}
\includegraphics[width=2.0\columnwidth]{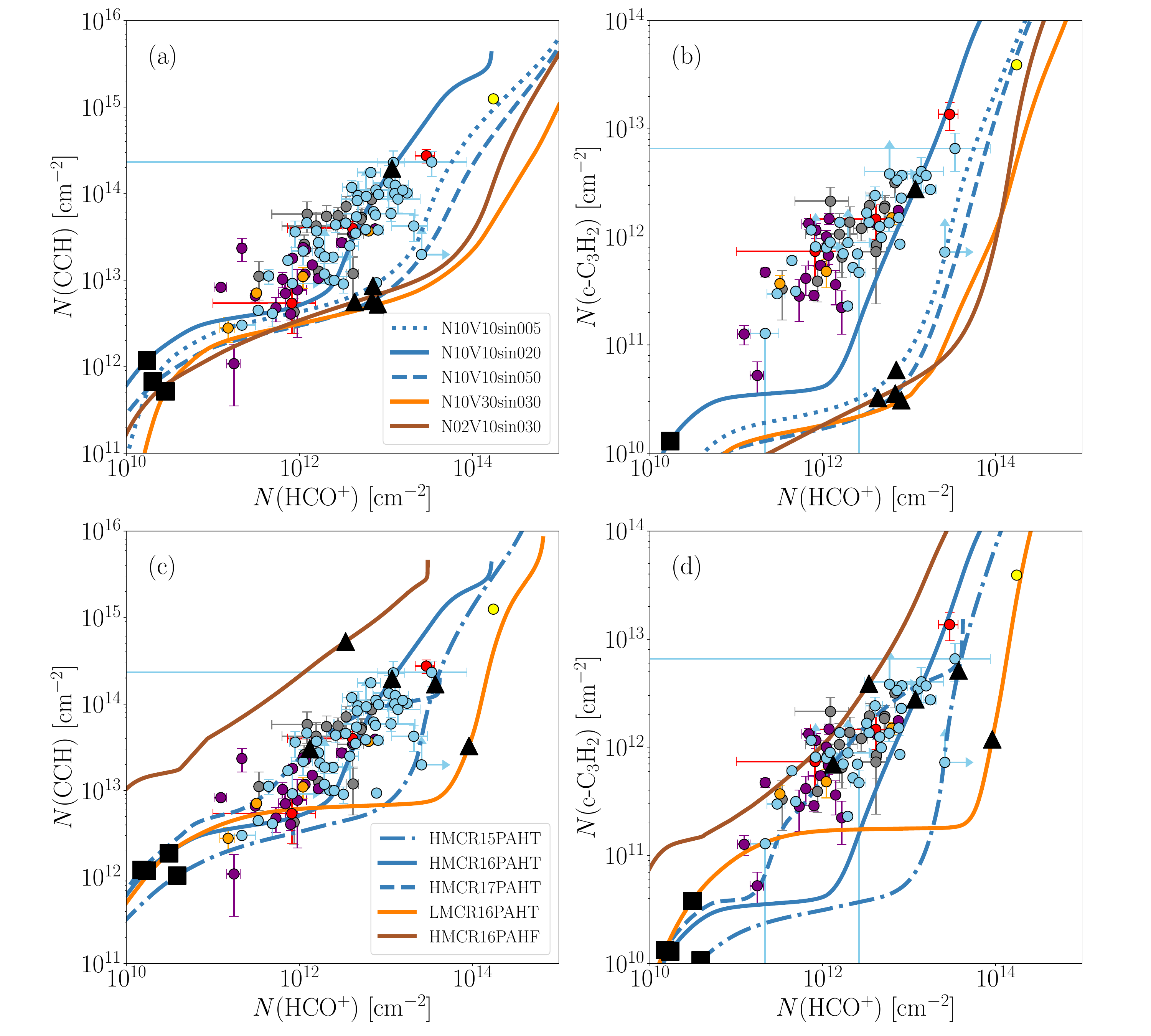}
\caption{\textrm{(a), (b) Comparison of the column densities of HCO$^+$ and carbon-chain molecules between observations (circles) and our models (lines). Dotted, solid, and dashed blue lines depict the model of  N10V10sin005, N10V10sin020, and N10V10sin050, respectively. The orange solid line is N10V30sin030, and the brown solid line is N02V50sin030. The black-filled squares and triangles represent the molecular column density of each model at $A_V^{\rm obs}=1$ and $5\,\mathrm{mag}$, respectively. The filled circles shows the observational values of \citet{1996A&A...307..237L} and \citet{2000A&A...358.1069L} (purple), \citet{2010A&A...520A..20G} and \citet{2011A&A...525A.116G} (gray), \citet{2011A&A...535A.103M} (yellow), \citet{2016PASJ...68....6A} (orange), \citet{2018A&A...610A..43R} (red), and \citet{2023A&A...670A.111K} (sky blue). (c), (d) Similar to panels (a) and (b), but the lines show the results of model N10V10sin020 with various chemistry parameters: HMCR16PAHT (solid blue line), HMCR15PAHT (dot-dashed blue line), HMCR17PAHT (dashed blue line), LMCR16PAHT (solid orange line), and HMCR16PAHF (solid brown line).}}
\label{fig:vsobs_CChain_physchem}
\end{figure*}

% \begin{figure}
% \includegraphics[width=\columnwidth]{figure/vsobs_CN_HCN_ete.pdf}
% \caption{The same as Figure \ref{fig:vsobs_CChain_physchem}, but for HCO$^+$, CN, and HCN. The purple circles represent the observational values from \citet{1996A&A...307..237L} and \citet{2001A&A...370..576L}. The other circles are the same as those in Figure \ref{fig:vsobs_CChain_physchem}.}
% \label{fig:vsobs_Nmol_phys}
% \end{figure}

% \begin{figure}
% \includegraphics[width=\columnwidth]{figure/vsobs_CN_HCN_chemparam_ete.pdf}
% \caption{The same as Figure \ref{fig:vsobs_CChain_physchem}, but for HCO$^+$, CN, and HCN.}
% \label{fig:vsobs_Nmol_chem}
% \end{figure}

\begin{figure*}
\includegraphics[width=2\columnwidth]{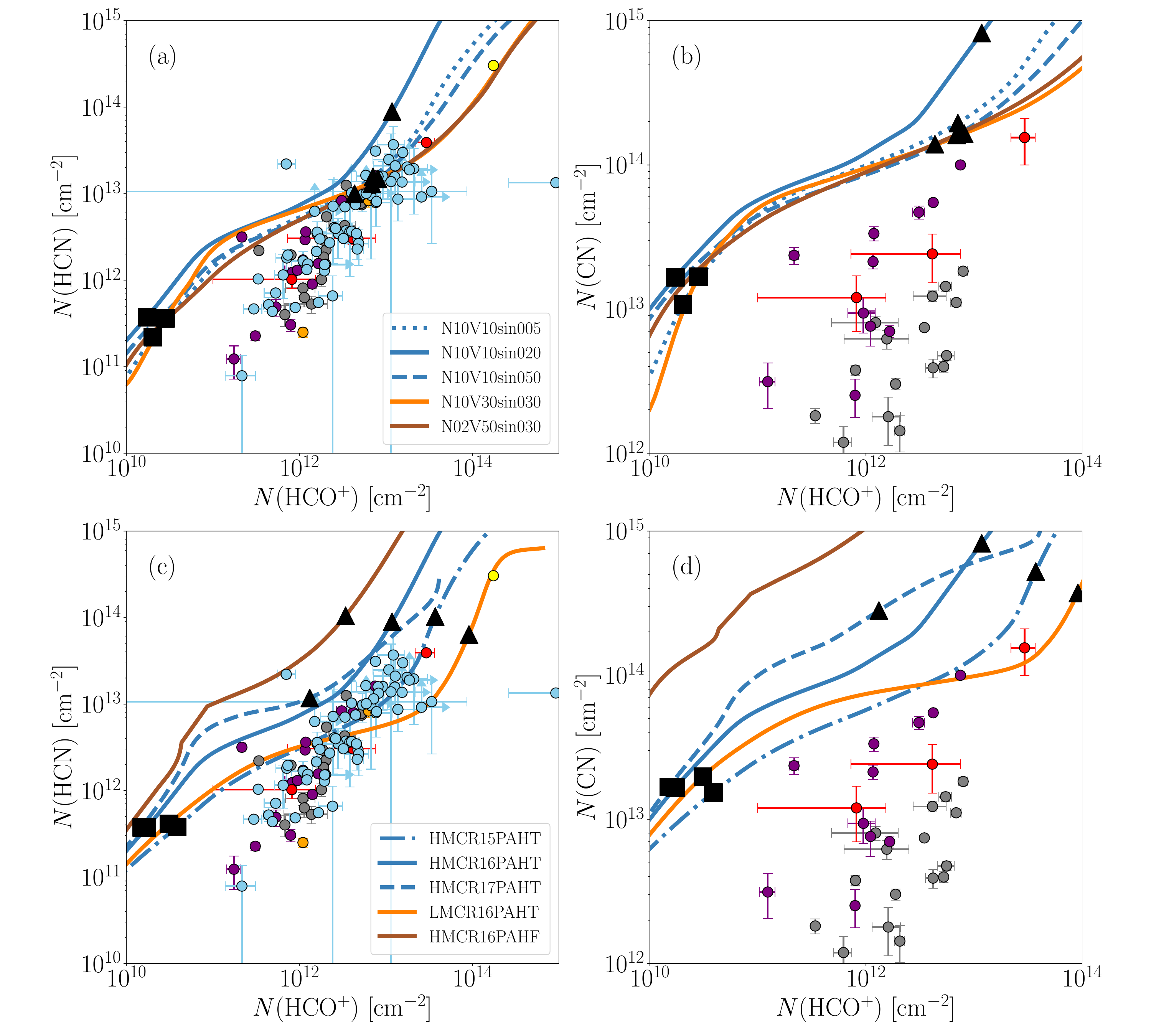}
\caption{\textrm{The same as Figure \ref{fig:vsobs_CChain_physchem}, but for HCO$^+$, CN, and HCN. The purple circles represent the observational values from \citet{1996A&A...307..237L} and \citet{2001A&A...370..576L}. The other circles are the same as those in Figure \ref{fig:vsobs_CChain_physchem}.}}
\label{fig:vsobs_Nmol_physchem}
\end{figure*}

\begin{figure}
\includegraphics[width=\columnwidth]{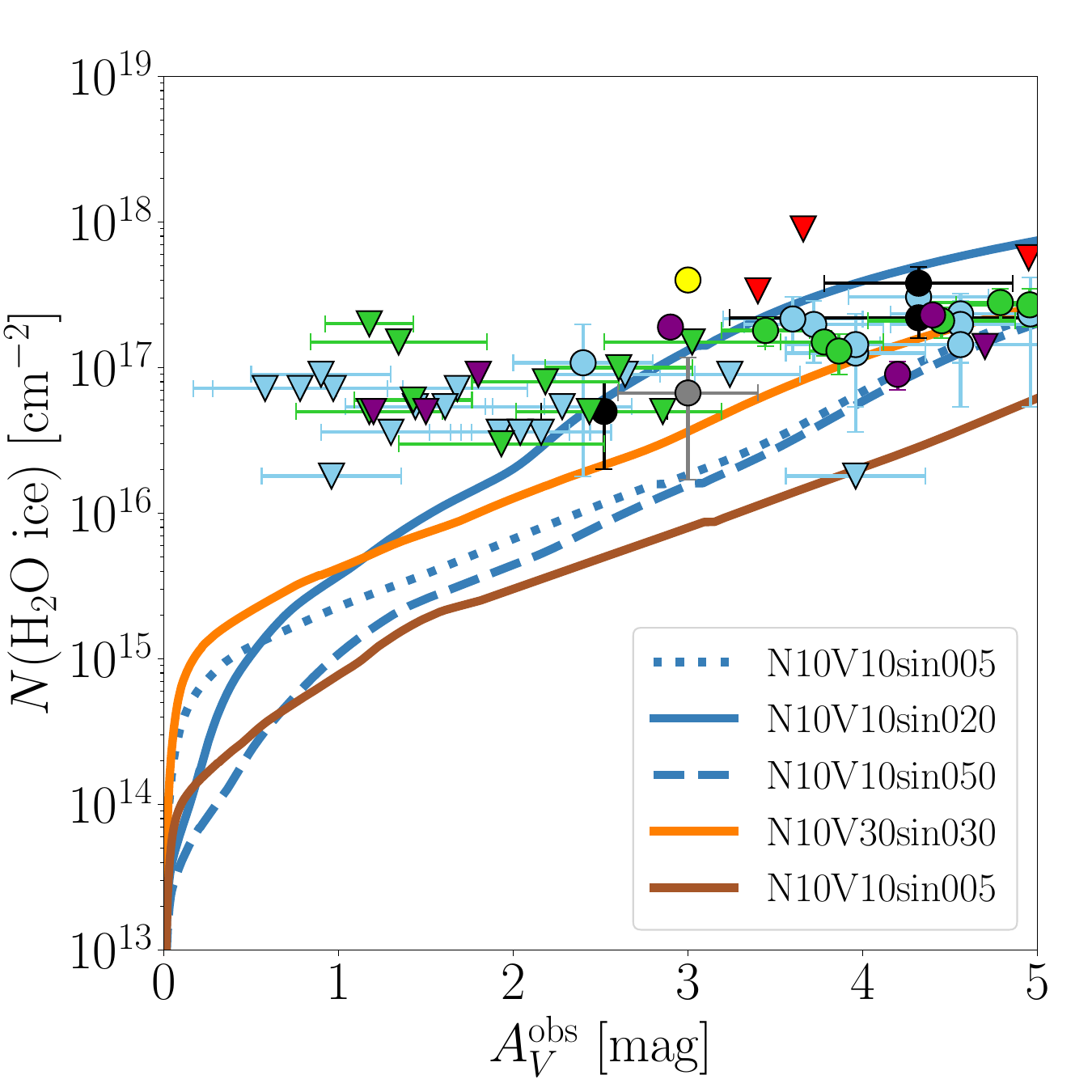}
\caption{Comparison of calculated column densities of water ice on dust grains with observations. Lines show the model results with various shock parameters; the style and color of the lines are the same as those of Figure \ref{fig:vsobs_CChain_physchem} (a). The filled circles and triangles represent the observational values of \citet{2013ApJ...777...73B} (green), \citet{2018A&A...610A...9G} (red), \citet{2000ApJS..128..603M} (sky blue), \citet{2007ApJ...655..332W} (blue), \citet{1999A&A...351..292T} (purple (Taurus) and yellow (Ophiuchus)), \citet{2013ApJ...774..102W} (black), \citet{2002MNRAS.330..837S} (gray), and references therein. The triangles show the upper limits.}
\label{fig:vsobs_waterice}
\end{figure}

\subsubsection{observations of gas-phase species}
\textrm{Panel (a) and (b) in Figure \ref{fig:vsobs_CChain_physchem} show} the column densities of CCH and c-C$_3$H$_2$ as a function of HCO$^+$ column density in the selected models with various shock parameters and those from observations of diffuse clouds. The results of our calculations show reasonable agreement with the observations, especially for CCH. The column density of carbon-chain molecules at $A_V^\mathrm{obs}\sim 5$ mag varies significantly among models (Section \ref{sec:result}), which reproduces the scatter of the observed column densities of carbon-chain molecules. As for c-C$_3$H$_2$, although some models underestimate its column densities, model N10V10sin020, which represents the most efficient compression, is consistent with the observations. It indicates that the diffuse clouds may have experienced efficient shock compression (but see the discussion on chemical parameters below). \citet{2010A&A...520A..20G} estimates the upper limit of gas density of the observed clouds to be $5\times 10^3$ cm$^{-3}$, which is also consistent with our models \textrm{(see panel (f) in Figure \ref{fig:physicalmodel_all})}. 

\textrm{Panel (c) and (d) in Figure \ref{fig:vsobs_CChain_physchem} compare} the observed column densities with the models of various chemical parameters. As discussed in Section \ref{sec:chemparams}, these parameters affect the ionization degree and the abundance of ions in the compression layers. Thus, the major difference between models appears in the column density of HCO$^+$. Compared with the fiducial case (HMCR16PAHT), HCO$^+$ column density is higher in the models with higher $\zeta_\mathrm{CR}$ and the low metal abundance, while it is lower in the model without PAHs. The variation in column densities due to these parameters overlaps with the observational values. The model with $\zeta_\mathrm{CR}=1\times 10^{-17}\,\mathrm{s^{-1}}$ shows lower column density of CCH ($\la 10^{14}$ cm$^{-2}$) than those with $\zeta_\mathrm{CR}=1\times 10^{-16}\,\mathrm{s^{-1}}$ and $1\times 10^{-15}\,\mathrm{s^{-1}}$ at $A_V^\mathrm{obs}\sim 5$ mag.

Our models with PAHs underestimate the c-C$_3$H$_2$ column density around $N$(HCO$^+$)$\sim 10^{11}-10^{12}$ cm$^{-2}$. One of the possible explanations is the uncertainty of the abundance and size of PAH. If the abundance of PAHs is smaller or their size is larger than assumed in our fiducial chemical model, the recombination rates between PAHs and cations are lower, and the blue and orange lines in Figure \ref{fig:vsobs_CChain_physchem} move toward model HMCR16PAHF shown in the brown line. While we assume the constant PAH abundance, it can decrease with time due to freeze-out onto grains \citep[e.g.][]{Hardegree-Ullman_2014}. Another possibility is the top-down chemistry. \citet{2022MNRAS.511.3832A} considered the formation of large hydrocarbons via the destruction of hydrogenated amorphous carbons by shock and showed that it can enhance the abundance of carbon-chain molecules within the timescale of shock propagation. For more quantitative studies of the top-down chemistry, however, we need to know the yield of carbon-chain production via the destruction of amorphous carbons as a function of shock velocity.

\textrm{Figure \ref{fig:vsobs_Nmol_physchem} shows} the comparison for CN and HCN. As described in Section \ref{sec:bfield_dep}, the dependence of their abundance and thus their column densities on shock parameters is not as strong as that of carbon-chain molecules. In the case of HCN, our models are consistent with the observations that show relatively high column density of HCN. However, our models tend to overestimate HCN column density especially at low $A_V$. 
The agreement is slightly better in the low metal abundance model (orange line in Figure \ref{fig:vsobs_Nmol_physchem}). As for CN, the observed column density is lower than that of our models. 

CN is formed via neutral-neutral reactions between hydrocarbon molecules and N atoms, and HCN formation is related to CN. Hence one way to suppress CN and HCN formation is to reduce the abundance of N atom in the gas phase. \textrm{Indeed, we confirmed that CN and HCN abundances decrease if we manually lower the initial N atom abundance in our calculations. In reality, N atom might be converted to other major N reservoirs, N$_2$, NH$_3$ ice, NH$_3$ salt, or unknown N reservoir, before it reacts with hydrocarbons to form CN and HCN.} 
In the fiducial model, N$_2$ reaches its peak abundance at $A_V\simeq 2$ mag  (Figure \ref{fig:N10V10sin020_mainspecies}), while it is mainly destroyed by photodissociation at $A_V\sim 1$ mag. N$_2$ can thus be more abundant if the self-shielding effect is enhanced. Multiple shock compression could enhance N$_2$ abundance. When a shock wave propagates perpendicular to the magnetic field, the post-shock density cannot be high enough for molecular cloud formation \citep[e.g.][]{Inoue_2009}, but N$_2$ is formed to some extent. If another shock wave arrives, N$_2$ can be abundant in the post-shock region due to the existence of N$_2$ in the pre-shock gas. NH$_3$ ice or its salt could accumulate in the solid phase in multiple shocks, as well. Multi-dimensional structure of the compression layer could also enhance the efficiency of the N$_2$ self-shielding, since N$_2$ formed in dense regions can be re-distributed in low density regions by turbulent mixing. 

Another solution for the disagreement is to increase HCO$^+$ abundance at $A_V\la 1$ mag. For example, \citet{2009A&A...495..847G,2014A&A...570A..27G} argued that decaying turbulent energy can be used to activate the endothermic chemical reaction to enhance the abundance of CH$^+$, which then reacts with O atom to form HCO$^+$. 
\citet{2022ApJ...926..190R} argued that 
the multi-phase structure of ISM can enhance the HCO$^+$ column density at $A_V\sim 1$ mag by considering thermally unstable regions with higher temperature than typical cold neutral medium \citep[see also][]{2020ApJ...903..142G}.
%While the present work considers the cloud formation by a single shock compression, interstellar gas can be swept by multiple shocks before a molecular cloud is formed 
%shock compressions occur about once per 1 Myr in our galactic disk
%\citep[e.g.][]{1977ApJ...218..148M,2015A&A...580A..49I,Hennebelle_2019}. This means that another shock compression 
Multiple shocks can temporarily heat the compression layer, which could also help the formation of HCO$^+$. 
%Therefore, considering multiple compressions in the model can be another solution for the deficiency of HCO$^+$ in our models.
These effects need to be considered in future work.

\subsubsection{observations of water ice}

Figure \ref{fig:vsobs_waterice} displays the column density of water ice in our models of various shock parameters with the chemical parameters of HMCR16PAHT. As expected from the water ice abundance described in Section \ref{sec:result}, the derived column density depends on shock parameters. This is due to differences in the post-shock density and the timescale of gas accumulation among models. N10V10sin020 and N10V30sin030 show relatively higher column density of water ice than other models due to high post-shock densities ($n_\mathrm{H}> 10^3\,\mathrm{cm^{-3}}$). The shorter timescale of accumulation in model N10V30sin030 has negative effects on water ice formation, as described in Section \ref{sec:parameter_velocity}. However, the water ice column density in N10V30sin030 is still higher than that of N10V10sin005 and N10V10sin050, whose post-shock densities are lower. This means that the effect of post-shock density is stronger than that of accumulation timescale at least within our parameter range. 

The results of the observations of water ice toward some dark clouds are also plotted in Figure \ref{fig:vsobs_waterice}. Since the present work focuses on diffuse and translucent regions, we choose the observational data at $A_{\rm V} \le 5$ mag. \textrm{As for the observational data in L183 (black circles), we evaluated $A_V$ from the optical depth of the 9.7 $\mathrm{\mu m}$ silicate feature $\tau_{9.7}$, i.e. $A_V=18\,\tau_{9.7}$ \citep{2013ApJ...774..102W}.} Although many of the data are upper limits or have large uncertainty, our models show reasonable agreement with them. Especially, N10V10sin020 and N10V30sin030, which have relatively high post-shock density, are consistent with observational values at $A_{\rm V}\ga 2$ mag. This indicates that the observed regions have moderately high densities ($n\ga 10^3\,\mathrm{cm^{-3}}$). 
%The black circles in Figure \ref{fig:vsobs_waterice} indicate the observed values toward L183, in which the threshold extinction for water ice detection is \textrm{$A_V\approx 2.7$ mag \citep{2013ApJ...774..102W}}. This threshold value is lower than that in Taurus (blue, sky blue, and purple points in Figure \ref{fig:vsobs_waterice}), $A_V\approx 3.2$ mag \citep{2001ApJ...547..872W}. The water ice column densities in N10V10sin020 are consistent with those of L183, while other models with lower post-shock densities are consistent with Taurus. Since high density can promote ice formation at lower visual extinction, this trend may indicate that density at around $A_V^\mathrm{obs}\la 5$ mag is higher in L183 than Taurus. 

One caution is that our model may overestimate water ice abundance at high visual extinction ($A_V^\mathrm{obs}>5$ mag). The predicted column density of water ice in our model is more than twice larger than the observed values in molecular clouds, e.g. Taurus \citep[e.g. see Figure 7 in][]{Boogert_2015}. As discussed in \citet{2010A&A...515A..66H}, this is due to too high oxygen atom abundance. In other words, there is an unknown reservoir of oxygen. Indeed, the carrier of several tens \% of the total oxygen abundance in molecular clouds is still unknown, while there is still a possibility that the abundance of water ice is underestimated in observations \citep{2021PhR...893....1O}.

\subsection{Future Prospect}
\label{sec:future_prospect}

Although we adopted the averaged structure of the compression layer to study its chemical evolution, the compression layer in the 3D MHD calculation has a wide range of density distribution. The probability density function of the total hydrogen number density of the compression layer has a lognormal-like shape. The volume filling factor of the region with mass-weighted density is $\sim 10\%$, while the total mass of the region with density larger than the mass-weighted density is $40-50\%$. Figure \ref{fig:vsobs_CChain_wm} shows the column densities of CCH and HCN as a function of HCO$^+$ column density in the fiducial model using the mass-weighted (orange lines) and volume-weighted densities (blue line). The model with mass-weighted density shows a higher CCH column density than the model with volume-weighted density, while the difference is smaller for HCN. If we fully consider the 3D structure, the actual molecular column densities should be between those of the mass-weighted case and the volume-weighted case. We will perform 3D MHD simulations coupled with chemical evolution in the subsequent work.

\section{Summary} \label{sec:conclusion}

We investigated the chemical evolution during the formation of molecular clouds by an interstellar shock wave. We conducted 3D MHD simulations of the converging flow of HI gas and investigated the effects of shock parameters, i.e. initial inclination of the interstellar magnetic field, initial shock velocity, and initial density. We analyzed the averaged physical structure of the compression layer, and derived the temporal variation of physical parameters of test particles that advect following the mean velocity field. Then we conducted the chemical network calculations along the mean flows. We also investigated how the molecular evolution depends on chemical parameters, i.e. abundance of refractory elements, cosmic-ray ionization rate, and the existence of PAHs. Our main findings are summarized as follows.

\begin{enumerate}
    \item The averaged post-shock gas density sensitively depends on the shock parameters, especially the inclination angle of the magnetic field to the conversing flow. This trend is consistent with the previous work \citep{2019ApJ...873....6I}.
    \item In classical pseudo-time-dependent models, the abundances of carbon-chain molecules, CN, and HCN reach the peak values when the C atom is the dominant carbon carrier. This trend is much less significant in our model, since the visual extinction of the compression layer is relatively low ($A_V \la 1$ mag) when C atom is abundant. Carbon chains increase at $A_V \ga 1$ mag, where the gas-phase C/O ratio increases due to water ice formation.
    \item The chemical desorption of hydrogenated nitrogen from the grain surface increases the abundance of NH$_n$ in the gas phase at $A_V\la 1$ mag. It also helps the formation of CN and HCN by the reactions between NH$_n$ and C$^+$.
    \item Molecular evolution in the cloud formation, i.e. the molecular abundances as a function of the visual extinction of the compression layer, depends on the gas density in the compression layer and the accumulation timescale, and thus on shock parameters. Carbon chains are especially enhanced in the models with high post-shock density and slow accumulation.
    \item The abundance of refractory elements, cosmic-ray ionization rate, and the existence of PAHs mainly affect the abundance of positively charged species (e.g. C$^+$, S$^+$, HCO$^+$, and H$_3^+$) and the ionization degree. Carbon chains are more abundant in the models with higher $\zeta_\mathrm{CR}$, in which more C$^+$ is produced by CO + He$^+$. 
    \item The molecular column densities along the direction of shock propagation are calculated from our models and compared with the observations of diffuse and translucent molecular clouds. The column densities of HCO$^+$ and carbon-chain molecules show reasonable agreement with observations. The variation of their column densities in the observations could be due to the difference in both shock parameters and chemical parameters. The column density of water ice is in reasonable agreement with the observations, as well. 
    \item The column densities of N-bearing molecules, i.e. CN and HCN, are overestimated in our models compared with the observed values. Their abundance could be reduced if N atom is converted to N$_2$, NH$_3$ ice, or NH$_3$ salt more efficiently by e.g. multiple shocks. Enhancement of HCO$^+$ at low $A_{\rm V}$ by extra heating provided by multiple shocks could also improve the agreement between our model with the observation. 
\end{enumerate}
While we focused on the averaged structure of the compression layers formed by a single shock wave, we will continue our study with more realistic 3D MHD simulations coupled with molecular evolution to investigate the effect of turbulence, the two-phase gas with density fluctuations, and multiple shocks.

\section*{software}
Athena++ \citep{2020ApJS..249....4S}, Numpy \citep{Harris_2020}, Matplotlib \citep{Hunter_2007}, Scipy \citep{2020SciPy-NMeth}

\section*{acknowledgments}
We thank Shu-ichiro Inutsuka, Tsuyoshi Inoue, Masato Kobayashi, Seiichi Sakamoto, Kohno Kotaro, Shota Notsu, German Molpeceres, Kanako Narita, and Shingo Hirano for constructive discussions, Wonju Kim, and Denise Riquelme for providing useful data on their observations. Numerical computations for the MHD simulations were carried out on Cray XC50 at Center for Computational Astrophysics, National Astronomical Observatory of Japan. 
This work is supported by JSPS KAKENHI grant Nos. 24KJ0901,
20H05847, \textrm{21K13967, JP21H04487, and JP22KK0043}.
Y.K. acknowledges the support by International Graduate Program for Excellence in Earth-Space Science (IGPEES) of the University of Tokyo.

%%%%%%%%%%%%%%%%%%%%%%%%%%%%%%%%%%%%%%%%%%%%%%%%%%
\section*{Data Availability}

Data will be made available upon reasonable request.



\bibliographystyle{mnras}
\bibliography{citation} % if your bibtex file is called example.bib

%\begin{thebibliography}{99}
%\end{thebibliography}

%%%%%%%%%%%%%%%%%%%%%%%%%%%%%%%%%%%%%%%%%%%%%%%%%%

%%%%%%%%%%%%%%%%% APPENDICES %%%%%%%%%%%%%%%%%%%%%

\appendix

\section{Chemical Evolution in HMCR16PAHF}
\label{app:bistability}

\begin{figure}
\includegraphics[width=0.9\columnwidth]{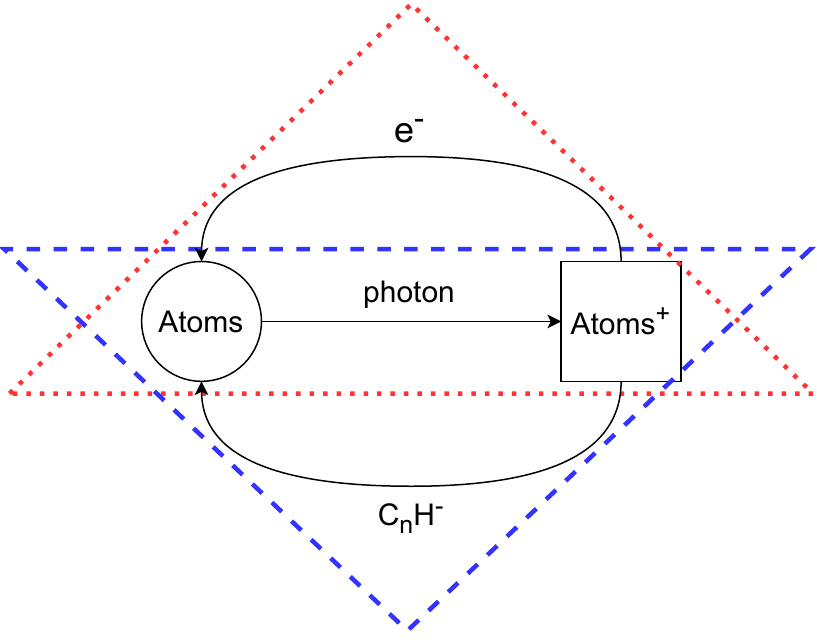}
\caption{The major chemical reactions for atomic ions in HMCR16PAHF. The reactions encircled by the red and blue triangles show the main reactions before and after the sudden decline of ionization degree, respectively.}
\label{fig:bistability}
\end{figure}

In model HMCR16PAHF i.e. the high-metal abundance model without PAHs, the molecular abundance and the ionization degree suddenly change at $A_V\simeq 1.9$ mag. This is due to the change in reactants of major charge carriers, atomic ions such as S$^+$ and Si$^+$, the abundance of which also sharply declines at  $A_V\simeq 1.9$ mag. Figure \ref{fig:bistability} displays the major reactions for the atomic ions. Before the decline, they mainly react with electrons. The typical rate coefficient for the radiative recombination is about $10^{-11}\,\mathrm{cm^3\,s^{-1}}$. As the C/O ratio increases in the gas phase, long carbon-chain molecules ($\mathrm{C_nH}$) become abundant, as explained in Section \ref{sec:fiducial}. They react with electrons to form anions ($\mathrm{C_nH^-}$), which become the major reactant with atomic ions:
\begin{equation}
\label{equ:recomb_metal_ani}
    \mathrm{C_nH^-} + \ce{M^+} \rightarrow \mathrm{C_nH} + \ce{M},
\end{equation}
where M represents the atoms that contribute to the ionization degree. The typical rate coefficient for this reaction is about $10^{-7}\,\mathrm{cm^3\,s^{-1}}$ \citep{Harada_2008}, which is orders of magnitude higher than that of radiative recombination. The ionization degree therefore decreases when the reaction (\ref{equ:recomb_metal_ani}) becomes effective. In addition, carbon-chain cations are mainly destroyed by dissociative recombination with an electron, which becomes less effective after the electron abundance decreases. This helps carbon-chain molecules and anions become more abundant and the ionization degree further decreases. We note that the major cation remains the same, i.e. S$^+$ and Si$^+$, before and after the decline of ionization degree.

CO also shows a temporal decrease and its chemistry changes at $A_V\simeq 1.9$ mag in model HMCR16PAHF. In the fiducial model with PAHs (HMCR16PAHT), CO is mainly formed by two neutral-neutral reactions: OH + C and C$_2$ + O. Formation of OH formation is initiated by the reaction between H$_3^+$ and O atom. Without PAHs, the electron abundance is high, which decreases the abundances of  H$_3^+$ and OH. Thus the reaction of C$_2$ + O dominates in CO formation in model HMCR16PAHF. As $A_V$ increases O atom starts to deplete from the gas phase, which makes the CO formation less efficient. At $A_V\simeq 1.9$ mag, carbon-chain molecules become abundant, and CO abundance also suddenly increases by reactions between the carbon-chain molecules and O atoms. 

\section{Mass-weighted vs. Volume-weighted Density}

\begin{figure}
\includegraphics[width=\columnwidth]{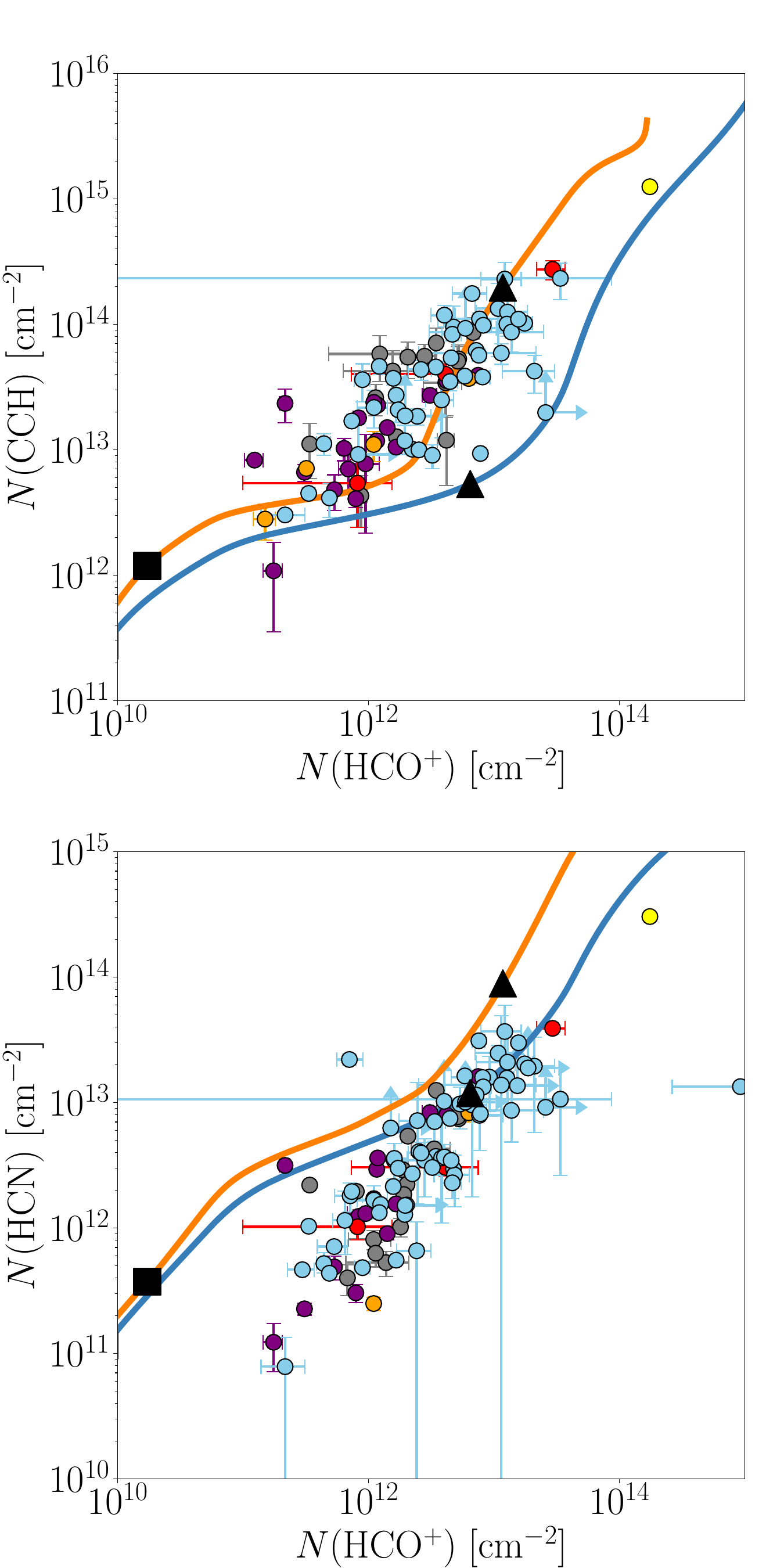}
\caption{Molecular column densities in the fiducial model using the mass-weighted (orange line) and volume-weighted gas density (blue line). Other details are the same as in Figure \ref{fig:vsobs_CChain_physchem}.}
\label{fig:vsobs_CChain_wm}
\end{figure}

\textrm{Figure \ref{fig:vsobs_CChain_wm} shows the column densities of CCH, HCN, and HCO$^+$ of N10V10sin020 with mass-weighted and volume-weighted density, which are discussed in Section \ref{sec:future_prospect}.}

%%%%%%%%%%%%%%%%%%%%%%%%%%%%%%%%%%%%%%%%%%%%%%%%%%

% Don't change these lines
\bsp	% typesetting comment
\label{lastpage}
\end{document}